\documentclass[global,twocolumn]{svjour} 
\usepackage{calc,color}
\usepackage{graphicx}
\usepackage[latin1]{inputenc}
\newcommand{\linbo}{$\rm LiNbO_3$}
\newcommand{\lisev}{~$\rm ^7LiNbO_3$:Fe}
\newcommand{\apbgraph}[1]{\includegraphics[width=\columnwidth]{#1}}
\newcommand{\href}[2]{{\tt\small #2}}
\title{The photo-neutronrefractive effect} 
\author{Martin Fally}
\institute{Institut f\"ur Experimentalphysik, Universit\"at Wien, Strudlhofgasse 4, A-1090 Wien, Tel.: +43-1-4277-52708, Fax.: +43-1-4277-9511, \email{Martin.Fally@exp.univie.ac.at}}
\date{\today}
\begin{document} 
\maketitle    
    \begin{abstract} 
Since its discovery in 1966, the photorefractive effect, i.e. the change of the refractive index upon illumination with light, has been studied extensively in various materials and has turned out to play a key role in modern optical technologies like photonics.

This article focuses on substances that change their refractive index for neutrons when irradiated with light. In analogy to light optics, we call them photo-neutronrefractive. After a short introduction to the relevant concepts of neutron optics, two materials exhibiting that effect, a photopolymer and an electrooptic crystal, are presented. Further, we discuss the the progress made concerning the development of creating light-induced gratings for neutron diffraction, which culminated in the setup of an interferometer for cold neutrons. Experiments performed on  photo-neutronrefractive materials are surveyed and the variety of corresponding results obtained is presented, including a discussion of their impact on material science, neutron optics, and the foundations of physics.    

\vspace*{10pt}

\end{abstract} 
      \section{Introduction \label{sec:introduction}}
When the subject of photorefractive effects began with the discovery of light-induced refractive-index in\-homo\-geneities in li\-thium niobate \cite{Ashkin-apl66}, neutron optics had already been established for more than 20 years as a by-product of the intense work with the Man\-hattan\--pro\-ject \cite{Anderson-pr46,Fermi-pr47,Zinn-pr47,Goldberger-pr47}. Both of the fields have evolved independently from each other into important branches of sciences and industry: the increasing importance of storing and quickly transferring large amounts of data resulted in a boom of designing optical elements (``Optical Information Processing, All Optical Networking"), making use of light-induced refractive-index changes \cite{Okamoto-or99}. Neutron optics, on the other hand, laid the foundations for a better understanding of quantum phenomena, which in turn was a necessary precondition to develop and make use of nanostructured materials or even quantum computers.
Among the highlights, one could mention the realisation of a commercially available holographic data-storage device \cite{Woike-pat00} and the successful setup and operation of a perfect-crystal neutron interferometer \cite{Rauch-pla74}, which marked a milestone in neutron optics and led to a resurgence of activity in this field.

Ten years ago those dynamic areas were linked by an experiment in which cold neutrons were diffracted from a grating created by a spatially inhomogeneous illumination of doped polymethyl\-methacrylate (PMMA). A typical holographic two-wave mixing setup was used to record a refractive-index pattern, a grating, in PMMA which was reconstructed not only with light, as usual, but also with neutrons \cite{Rupp-prl90}. Evidently, the illumination induced refractive-index changes for both, light \emph{and} neutrons! In analogy to light optics we call this behaviour the photo-neutronrefractive effect.
The basic procedure is characterised by two steps: the preparation of the gratings by means of nonlinear optics and the diffraction of neutrons from that grating. 
This enables us to achieve either of our aims: to extract information about the sample, to use the gratings for neutron-optical purposes, or to get insight into fundamental properties of the neutron itself. It is worth mentioning that in this way nonlinear light optics (photorefraction, electrooptics) and linear neutron optics are combined.
Subsequently, great effort was devoted to improving the technique of sample preparation, to understanding underlying physics and chemistry, to enhance the photo-neutronrefractive effect, to develop and design optical elements, and to use those devices in tackling fundamental problems in physics. A number of Master and Ph.D. theses have already dealt with several aspects related to these topics \cite{Kopietz-85,Matull-91,Kohlbrecher-92,Danckert-94,Breer-95,Grossmann-95,Kuhnt-97,Schellhorn-98,Pruner-98,Havermeyer-00}. The major results, mostly unpublished though quite interesting, will also be summarised in this review.

The article is organised as follows: Starting with a concise explanation of the relevant concepts in neutron optics, electrooptics and photorefraction, as well as diffraction phenomena, we introduce PMMA and the electrooptic crystal \linbo~ as photo\--neutron\-refractive materials. Further, we discuss the preparation of the gratings by means of the holographic two-wave mixing technique, which is a crucial point for manufacturing neutron-optical elements like mirrors, beam splitters, lenses or interferometers. The main part is concerned with the neutron diffraction experiments performed on deuterated PMMA (d-PMMA). We show that this type of experiments can be useful when studying the polymerisation process itself, when serving simply as a neutron-optical element, or when probing fundamental properties of the neutron. The latter is in particular true of electro neutron-optic \linbo, where the diffracted neutrons are inherently exposed to extremely high electric fields due to the light-induced charge transport. The corresponding experiments already conducted are presented together with those that still need to be done.
Finally, we discuss the future perspectives of photo-neutronrefractive materials and their possible applications. 
\section{Basic concepts}
This part will provide the necessary prerequisites in photorefraction, holography, neutron optics and diffraction physics to be able to understand the experiments and the obtained results. At first, we will present the technique for preparing the gratings, introduce the equation of motion for neutron diffraction, define the neutron-optical potential and the neutronrefractive index, and finally discuss the relevant terms of the neutron-optical potential, which are modulated by inhomogeneous illumination with light.
      \subsection[Holographic Technique]{Preparation of the gratings\label{sec:holography}}
Soon after the discovery of photorefraction \cite{Ashkin-apl66}, the technological importance of the effect became clear, e.g. that such materials can be used for information storage and as holographic memories \cite{Chen-apl68}. The big advantage over materials changing their absorption, like standard photographic films, is that intensity losses do not occur, and thus the whole volume rather than the surface may be used for data storage. Important consequences when utilising thick volume phase gratings in diffraction experiments are that a sharp Bragg condition must be obeyed and that multiple scattering effects must be taken into account, i.e. dynamical diffraction theory has to be considered. Historically, the latter was originally developed for X-rays by Darwin, Ewald and von Laue at the beginning of the 20th century and extended in several review articles (see e.g.  \cite{Batterman-rmp64}). When lasers became available and the technique of holography had experienced its first climax, Kogelnik reinvestigated the effects of coupled waves for light \cite{Kogelnik-Bell69}. Finally, when crystals of highest quality and thickness could be produced as a consequence of semiconductor technology, Rauch and Petrascheck performed this task for neutrons \cite{Dachs-78}. We will summarise the results of this theory in so far as they are necessary to interpret our experiments correctly.
\subsubsection{Recording holographic gratings}
\begin{figure}
\begin{center}
    \apbgraph{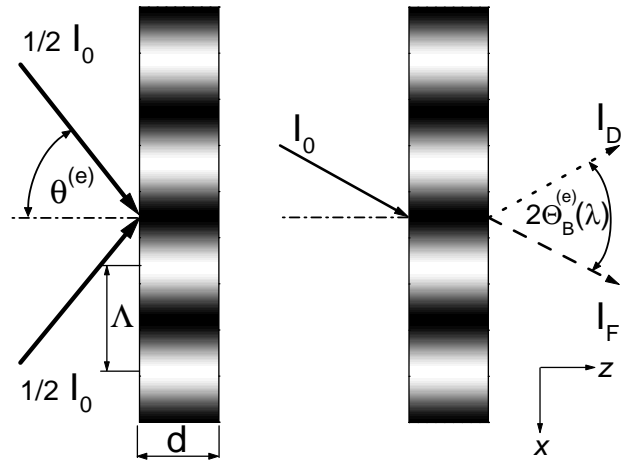}
  \caption{\label{fig:1} Sketch of the setup for the preparation of light-induced refractive-index gratings (hologram recording) and the reconstruction with light or neutrons.}
\end{center}
      \end{figure}

As a first step, we will discuss the typical setup for the preparation of light-induced refractive-index gratings, which is sketched in Figure  \ref{fig:1}. 
Two coherent plane light waves interfere in the photo-neutronrefractive material (two-wave mixing). In the simplest case with waves of equal intensity and mutually parallel polarisation states, the resulting modulation of the light pattern is sinusoidal $\Delta I(x)\propto \cos{(Kx)}$ with a grating spacing 
      \begin{equation}
\Lambda=\frac{2\pi}{K}=\frac{\lambda}{2 \sin{(\theta^{(e)})}}.
      \end{equation}
Here $2\theta^{(e)}$ denotes the angle between the interfering beams in air and $\lambda$ the wavelength of light in vacuum. 
Angles in air are indicated by the superscript $(e)$, otherwise angles in the medium are meant.
Typical values for the grating spacing are 300 nm$<\Lambda<$ 2000 nm.
In general, this inhomogeneous illumination of the sample results in a spatially dependent refractive-index change which can be expanded in a Fourier series
      \begin{equation}\label{eq:dn-harm}
\Delta n(x)=\sum_j \Delta n_j \cos{\left(j Kx +\phi_j\right)}.
      \end{equation}
The actual pattern of course depends on the properties of the material and the mechanism of photorefraction. Usually in electrooptic crystals a linear response, i.e. $\Delta n(x)\propto \Delta I(x), \Delta n_{j>1}\equiv 0$, is obtained though it may be nonlocal ($\phi_1\not=0$). 
This is not always the case: e.g., the photorefractive effect in PMMA depends on illumination time and intensity and is strongly nonlinear as will be shown in section \ref{sec:2harm}.
      \subsubsection{Diffracting light and neutrons}
Diffraction from such refractive-index gratings which represent thick holograms is governed by the basic formulae of the dynamical diffraction theory. In our experiments we only deal with nearly lossless dielectric gratings in the transmission geometry. To gain information about the material parameters, light or neutrons are diffracted from those gratings in the vicinity of the Bragg angle $\Theta_B^{(e)}=\arcsin{[\lambda/(2\Lambda)]}$. In particular, the diffraction efficiency $\eta(\theta)=I_D/(I_D+I_F)$ is measured as a function of the deviation from $\Theta_B$, the so-called rocking-curve. $I_{D,F}$ are the diffracted and the forward diffracted intensities respectively. For a monochromatic plane wave in symmetric transmission geometry the diffraction efficiency is described by \cite{Kogelnik-Bell69}
      \begin{eqnarray}\label{eq:diffractionefficiency}
\eta(\theta,\nu)&=&\nu^2{\rm sinc}^2\left(\sqrt{\nu^2+\xi(\theta)^2}\right)\\
\label{eq:nu}
\nu&=&\frac{\pi\Delta n d}{\lambda\cos{(\Theta_B)}}\\
\label{eq:xi}
\xi(\theta)&=&\frac{K(\Theta_B-\theta) d}{2}=\frac{\pi (\Theta_B-\theta) d}{\Lambda}.
      \end{eqnarray}
The thickness of the grating, which in the ideal case is identical to the sample's thickness, is denoted by $d$. The parameters $\nu$ and $\xi$ contain the relevant material information. In particular, the light-induced refractive-index change $\Delta n$ can be probed and determined by such a measurement. Another useful measure is the integrated diffraction efficiency (i.e. integrated reflectivity times $\nu$)
      \begin{equation}\label{eq:BWGint}
I(\nu):=\frac{d}{\Lambda}\int_{-\infty}^{+\infty}\eta(\theta,\nu)d\theta=
\frac{\nu}{2}\int_0^{2\nu}J_0(s)ds,
      \end{equation}
with $J_0(s)$ the zeroth-order Bessel function. The practical importance of $I(\nu)$ lies in the fact that this quantity is independent on the lateral divergence of the beam and that it can be accessed experimentally by performing rocking-curves (see section \ref{sec:timeev}). 

\subsection{Neutron optics \label{sec:neutron optics}}
\subsubsection{Definition of the neutronrefractive index}
The equation of motion for a field of non-relativistic particles, e.g. cold neutrons with a kinetic energy of $0.4$K$<E/k_B<40$K or a wavelength $5$\AA$<\lambda<50$\AA, is Schrö\-din\-ger's equation. We restrict our considerations to \emph{coherent elastic scattering} (neutron optics) in condensed matter. Then the coherent wave $\Psi$ and the coherent scattering are described by a one-body Schrödinger equation with the (time-independent) neutron-optical potential $V(\vec{x})$ \cite{Sears-89} and the energy eigenvalues $E$:
      \begin{eqnarray}\label{eq:schroedinger}
{\cal H}\Psi(\vec{x})=E \Psi(\vec{x})\\
\label{eq:potential}
{\cal H}=-\frac{\hbar^2}{2m}\nabla^2+V(\vec{x}).
      \end{eqnarray}
For a general treatment of neutron optics see e.g. Refs. \cite{Sears-89,Klein-rpp83}.
Inserting (\ref{eq:potential}) in (\ref{eq:schroedinger}) leads to a Helmholtz type equation, with $k_0$ being the vacuum wave vector, if we properly define the refractive index for neutrons $n_N$:
      \begin{eqnarray}\label{eq:helmholtz}
\left[\vec{\nabla}^2+(n_N(\vec{x})k_0)^2\right]\Psi(\vec{x})=0 \\
\label{eq:nindex}
n_N(\vec{x})=\sqrt{1-\frac{V(\vec{x})}{E}}.
      \end{eqnarray}
It is evident that any change in the potential of matter results in a refractive-index change for neutrons. In our particular context the main task is to modify $V(\vec{x})$ by illumination with light, i.e. to observe a photo-neutronrefractive effect. Therefore it is necessary to discuss the relevant terms of the neutron-optical potential.
Interactions between the neutron and condensed matter may be classified pragmatically into three groups according to their magnitude \cite{Shull-taca67,Sears-prep86}: The strong interaction dominates in non-magnetic materials (``bread and butter''-effect) whereas the electromagnetic neutron-atom interaction is at least 2 to 3 orders smaller (``nuisance''-effects). The latter, however, is important if high electric fields are applied and thus must be taken into account (see section \ref{sec:enoc}).
      \subsubsection{Nuclear contribution to the potential}
As a consequence of the strong interaction, the scattering amplitude of cold neutrons is proportional to a constant and independent of the scattering angle in first order approximation, because of the extremely short interaction length. Therefore the nuclear potential is replaced by the Fermi pseudo-potential 
      \begin{equation}\label{eq:fermi0}
v_N(\vec{x})=\frac{2\pi\hbar^2}{m}b\delta(\vec{x}-\vec{y}),
      \end{equation}
with $m$ the neutron rest mass and $b$ the coherent nuclear scattering length, which in general is spin-dependent. Dealing with bulk matter, we are interested in the macroscopic optical potential $V_N(\vec{x})$, which is simply given by a series of Fermi pseudo-potentials
      \begin{equation}\label{eq:fermi}
V_N(\vec{x})=\frac{2\pi\hbar^2}{m}\sum_j b_j\delta(\vec{x}-\vec{y}_j)=\frac{2\pi\hbar^2}{m}
b^D(\vec{x}).
      \end{equation}
Summation is performed over the different nuclei $j$ with scattering length $b_j$ at the corresponding sites $\vec{y}_j$. $b^D(\vec{x})$ denotes the so-called coherent scattering length density. Assuming that the internal degrees of freedom of the atoms are statistically independent of their positions, (\ref{eq:fermi}) can finally be simplified as
      \begin{equation}\label{eq:nopt}
V_N(\vec{x})=\frac{2\pi\hbar^2}{m} \overline{b}\rho(\vec{x}).
      \end{equation}
Here $\overline{b}$ is the mean bound scattering length averaged over a unit, e.g. a unit cell in a regular crystal or a polymer unit, and $\rho(\vec{x})$ is the number density.

The photo-neutronrefractive effect in PMMA is based on the fact that the photopolymer has a higher number density than the monomer MMA. By illuminating the photosensitised sample with a sinusoidal light pattern, we also modulate the number density $\rho(\vec{x})=\overline{\rho}
+\Delta\rho(\vec{x})$ sinusoidally. Thus the neutron-optical potential reads:
      \begin{equation}\label{eq:densvar}
V_N(\vec{x})=V_N+\Delta V_N(\vec{x})=\frac{2\pi\hbar^2}{m}
\left(\overline{b}^D+\overline{b}\Delta\rho(\vec{x})\right),
      \end{equation}
with the mean scattering length density per unit volume $\overline{b}^D$.
      \subsubsection{Contribution to the potential due to an electric field}
In this section we will consider the influence of a (static) electric field $\vec{E}(\vec{x})$ on the neutron-optical potential. From (\ref{eq:fermi}) it is evident that under the application of an electric field the nuclear contribution of the potential $V_N$ can be influenced by either changing at least one of the scattering lengths or at least one of the partial number densities $\delta(\vec{x}-\vec{y}_j)=\rho_j(\vec{x})$. Changes of the scattering length could be established by an influence of the electric field on the nuclear polarisability. Here, we discuss only its influence on the number density which is larger by orders of magnitude. 
Let us assume an electrooptic crystal which thus is also piezoelectric by symmetry. Application of an electric field $\vec{E}(\vec{x})$ will hence lead to a strained crystal, i.e. to a density variation $\Delta\rho(\vec{x})$. Again (\ref{eq:densvar}) is valid. The magnitude of the density modulation then depends on the symmetry as well as the values of the compliance and the piezoelectric tensor \cite{Havermeyer-apb99}.

Moreover, a neutron moving in an electric field $\vec{E}(\vec{x})$ with velocity $\vec{v}$ gives rise to an additional contribution to the potential (Schwinger term, spin-orbit coupling, Aharonov-Casher effect) due to its magnetic dipole moment $\vec{\mu}$ \cite{Schwinger-pr48}. 
Its canonical momentum then is $\vec{p}=m\vec{v}+(\vec{\mu}\times\vec{E})/c$.
In the standard derivations of this term $\vec{E}(\vec{x})$ usually denotes the electric field of the atoms or an electric field in vacuum. The Schwinger term reads:
\begin{equation}\label{eq:Schwinger}
V_S(\vec{x})=-\frac{\hbar}{mc}\vec{\mu}\cdot
\left[\vec{E}(\vec{x})\times\vec{k}\right].
\end{equation}
However, when instead an electric field is applied on a crystal, its static dielectric constant $\varepsilon$ must be taken into account the potential $V_S$ therefore is enhanced by an additional factor of $\varepsilon$.

Further terms in the expansion of the potential which are linear in the electric field are the Foldy contribution \cite{Foldy-pr51,Foldy-pr52}
      \begin{equation}\label{eq:Foldy}
V_F(\vec{x})=-\frac{\hbar|\vec{\mu}|}{2mc}\left[\vec{\nabla}\cdot\vec{E}(\vec{x})\right],
      \end{equation}
and the tiny part (``trivial interaction'' according to Shull's classification \cite{Shull-taca67}) due to the electric dipole moment $\vec{d}$ (EDM) of the neutron 
      \begin{equation}\label{eq:EDM}
V_{EDM}=\vec{d}\cdot\vec{E}(\vec{x}).
      \end{equation}
\subsubsection{Estimation of the neutronrefractive index}
Finally, we estimate the neutronrefractive index for the materials which will be discussed in next sections. If we take into account only the leading terms in the potential, i.e. employing  (\ref{eq:fermi}), the neutronrefractive index is
      \begin{equation}\label{eq:nref}
n_N=\sqrt{1-\frac{\lambda^2}{\pi} \overline{b}^D}\approx 1-\frac{\lambda^2}{2\pi}\overline{b}^D.
      \end{equation}
Note the quadratic dependence on the wavelength $\lambda$. In Table  \ref{tab:nref} the mean coherent scattering length densities and the refractive indices are tabulated for PMMA, d-PMMA and $\rm ^7LiNbO_3$, for two typical cold neutron wavelengths.
      \begin{table}
  \centering
\begin{tabular}{|l|r@{.}l|r@{.}l|r@{.}l|} \hline
 &\multicolumn{2}{|c|}{} & \multicolumn{2}{|c}{$\lambda=1$ nm} & \multicolumn{2}{c|}{$\lambda=3$ nm}\\
& \multicolumn{2}{|c|}{$\overline{b}^D$ [$10^{14}$/m$^2$]}& \multicolumn{4}{|c|}{$n-1 [10^{-5}$]}\\ \hline
  PMMA & 1&03 & -1&64 & -14&7 \\
  d-PMMA & 6&26 & -9&96& -90&0 \\ 
$\rm ^7LiNbO_3$ & 4&04  & -6&43 & -57&9 \\ \hline
\end{tabular}
\caption{Coherent scattering length densities and neutronrefractive indices for PMMA, d-PMMA and $^7$\linbo~ respectively. \label{tab:nref}}
      \end{table}
For most of the experiments, except the very first ones, d-PMMA was used. The large incoherent scattering cross section of H-atoms was an obstacle for high diffraction efficiencies. The reason for using monoisotopic $\rm ^7Li$ is to avoid  the high absorption of $\rm ^6Li$ which is contained in natural \linbo.
At this point it is worth emphasising that in photo-neutronrefractive materials we are dealing with \textbf{changes of the refractive index}. This is quite a challenging task for neutrons. 
\section{Materials}
So far two different types of photo-neutron\-refractive effects have been realised: changes of the optical potential $\Delta V\propto\Delta\rho$ resulting from chemooptics in photopolymers and $\Delta V\propto E$ resulting from electro (neutron)optics in crystals.
Here, we will discuss the basic mechanisms of photo-(neutron)\-refraction in such materials, represented by PMMA and LiNbO$_3$ re\-spec\-tively.
\subsection{Poly(methylmethacrylate)\label{sec:PMMA}}
The polymer PMMA is well known in everyday life as plexiglass. The basis for the existence of a photo-neutronrefractive effect in poly(methylmethacrylate) (PMMA) is the large difference of number densities for the monomer MMA and the polymer PMMA, $\rho_{MMA}:\rho_{PMMA}=0.94$ g/cm$^3:1.18$ g/cm$^3$. A photo-induced polymerisation then allows us to modulate the density and hence the neutronrefractive index by illumination. Moreover, the mechanical and, in particular, the excellent optical properties made PMMA the favorite candidate not only for fundamental studies but also for potential technical applications.
\subsubsection{Pre-polymerisation process}
To polymerise the monomer MMA, a $\rm (C=C)$ double bond  must be split. This is established by free radicals which are created by a thermoinitiator. Typically $\alpha,\alpha'$-Azo-isobutyronitrile (AIBN) \cite{Kopietz-85,Matull-91,Kuhnt-97,Pruner-98} was used. At elevated temperatures AIBN forms two radicals which react with MMA and start a chain reaction resulting in the polymer. At temperatures below 330 K two major mechanisms lead to the termination of the polymerisation: either by combination, i.e. by two radicals forming a bond, or by disproportionation, i.e. by one of the two radical chains transferring an H-atom and forming a double bond, thus stopping the growth of both chains. As a consequence of the low temperature polymerisation ($T\approx325$ K), part of residual monomers remain solved in the PMMA-matrix. 
\subsubsection{Light induced post-polymerisation}
Those monomers now served as reservoir to restart the polymerisation by illumination. To sensitise the material a photoinitiator had been added prior to pre-polymerisation. Depending on the application, photosensitive substances for the ultraviolet or visible spectral range were used. When irradiating the samples with light, the photosensitive component decomposes into free radicals which restart the polymerisation in the bright regions. This yields a density modulation and, according to  (\ref{eq:densvar}), a neutronrefractive-index change. Photoinitiators employed were 2,2-dimethoxy-1,2-diphenyl-ethanone (DMDPE) \cite{Havermeyer-00} or 2,2-dimethoxy-2-phenyl-acetophenone (DMPA) \cite{Havermeyer-apb01} which both are sensitive to ultraviolet light. The use of a short wavelength ($\lambda=351$ nm) was favourable as considerably smaller grating spacings could be reached. Typical recording intensities were in the range of several hundred W/cm$^2$. Exposure times between 2 and 60 seconds were employed. During the past few years this parameter has turned out to be an important quantity for the photorefractive response (see section \ref{sec:2harm}).
\subsubsection{The photorefractive mechanism in PMMA\label{sec:absmodel}}
The photosensitivity of the doped PMMA system becomes manifest primarily in light-induced absorption changes $\Delta\alpha(Q)$ \cite{Kopietz-85,Havermeyer-apb01}. PMMA/DMDPE shows the peak of the absorption band around $\lambda=350$ nm, which is rather convenient for the Ar-ion uv laser line. Upon illumination this band vanishes and shifts the fundamental absorption band edge by about 20 nm to lower energies. Following the kinetics of $\Delta\alpha$ two processes are involved: after an increase and after reaching its peak value, the light-induced absorption decreases to negative values. When stopping the illumination no absorption change is observed at all. Havermeyer et al. proposed a model attributing one of the processes to the decay of the photoinitiator into radicals and inert molecules \cite{Pruner-98,Havermeyer-apb01}. This mechanism, which is responsible for the polymerisation, leads to a permanent change of the refractive index: the photorefractive effect. The second process was assumed to stem from a light-induced termination for the radicals which linearly depended on the intensity of the light. The solution of their rate equation, which nicely described the experimentally obtained results, yielded for the exposure dependence of the light-induced absorption changes:
      \begin{equation}\label{eq:abs-model}
\Delta\alpha(Q)=a_1 \left[1-\exp{(-k_1Q)}\right]+a_2 \left[\exp{(-k_2Q)}-1\right],
      \end{equation}
with $Q=I_0t$ the exposure and $a_i,k_i$ being proportionality or rate constants respectively.
Via the Kramers-Kr\"onig relations the corresponding refractive-index change $\Delta n$ can be obtained. It should be mentioned that in this multicomponent system (monomer, oligomer, polymer) the polarisability $\Pi$ is made up simply by the weighted polarisabilities of the single components whereas this is not the case for the densities \cite{Panke-Mc73}. Hence, also the refractive-index change cannot simply be assembled from the refractive-index changes of its constituents.
\subsubsection{Experimental conditions}
To record gratings in PMMA the expanded beams of an Ar-ion laser operating at one of the uv-lines ($\lambda=333, 351, 363$ nm) were used.  Typical intensities ranged from 50 W/m$^2$ to 500 W/m$^2$. The samples were exposed to the interference pattern (see Figure \ref{fig:1}) only for a few seconds up to half a minute. Then the recording light was switched off and the temporal evolution of the gratings was followed in the dark by diffraction with neutrons and probe-light of low intensity (see section \ref{sec:timeev}f.). 
\subsection{\linbo\label{sec:linbo}}
\linbo was the first photorefractive material to be discovered \cite{Ashkin-apl66,Chen-apl68}. Its photorefraction is based on the excitation, migration and trapping of charges when illuminated by coherent light radiation. A space-charge density is building up, which, according to Poisson's equation, leads to a space-charge field $E_{sc}$ and via the electrooptic effect to a refractive-index change $\Delta n$ for light, see e.g.  \cite{Gunter-1-87}. When illuminating the photorefractive sample with a sinusoidal light pattern as described in section \ref{sec:holography}, the space-charge field in the stationary state is given by
\begin{equation}\label{eq:esc}
  E_{sc}(x)=-E_1 \cos{(Kx+\phi)}.
\end{equation}
Here $E_1$ is the magnitude of the effective electric field, i.e. the first coefficient in a Fourier series, which depends on the recording mechanism. In \linbo:Fe the photovoltaic effect is dominant and thus $E_0\approx E_{PV}$ and $\phi\approx 0$. Employing the linear electrooptic effect (Pockels-effect), the refractive-index change for light then equals
\begin{equation}\label{eq:eo}
  \Delta n_L(x)=-\frac{1}{2}n_L^3r E_{sc}(x),
\end{equation}
with $n_L$ the refractive index for light and $r$ the effective electrooptic coefficient.
\subsubsection{The electrooptic effect revisited}
It seems worth discussing the linear electrooptic effect more accurately already at this point. The electrooptic tensor $r_{ijk}$ can be understood as a coefficient in the expansion of the dielectric displacement $D_i$ with respect to the electric field $E_j$.
Considering in addition the elastic degrees of freedom, it is necessary to define which of the thermodynamic variables - stress $\tens{T}$ or strain $\tens{S}$ -  are to remain constant when performing the derivatives. In an experiment both of these cases can be realised. Keeping the strain constant results in the clamped linear electrooptic coefficient $\tens{r^S}$. The applied electric field changes the refractive index directly, i.e. by slightly modifying the electronic configuration. However, when keeping the stress constant, the free linear electrooptic coefficient $\tens{r^T}$ is measured. Here, in addition, a contribution via the piezoelectric coupling ($d_{ijk}$) in combination with the elastooptic effect ($p_{ijlm}$) must be considered $r_{ijk}^T=(r_{ijk}^\tens{S}+p_{ijlm}^Ed_{klm})$ (see e.g. Ref.  \cite{Jazbinsek-apb02}).
When experimentally realising the latter case, we arrive at
      \begin{equation}\label{eq:deltanl}
\Delta\!\left(\frac{1}{n_L^2}\right)_{ij}=r_{ijk}^\tens{T}E_k=
\left(r_{ijk}^\tens{S}+p_{ijlm}^Ed_{klm}\right)E_k
      \end{equation}
for the tensor of the optical indicatrix.
In \linbo~ $\tens{r^S}$ contributes about 90\% of the polarisability to $\tens{r^T}$. This is plausible as light is quite sensitive to electronic changes but much less to density variations.
\subsubsection{The electro neutron-optic effect}
Recalling section \ref{sec:neutron optics}, it becomes evident that in analogy to light electrooptics we can define the corresponding effect for neutrons: the \textit{electro neutron-optic effect}. Utilising  (\ref{eq:nindex}) and neglecting the tensorial character of the effect, the electro neutron-optic effect reads
      \begin{equation}\label{eq:deltann}
    \Delta\!\left(\frac{1}{n_N^2}\right)=\frac{2m}{(\hbar k)^2}\Delta V=r_N E(\vec{x}).
      \end{equation}
Consequently, we call the proportionality constant $r_N$ the \textit{electro neutron-optic coefficient} (ENOC).
By inserting  (\ref{eq:densvar}) to (\ref{eq:EDM}) into  (\ref{eq:deltann}) and comparing the corresponding terms, we can identify the following relations
      \begin{eqnarray}\label{eq:enocS}
r^S_N&=&\frac{2 \lambda |\vec{\mu}|}{hc}
\left(\pm\varepsilon+i\frac{\lambda}{2\Lambda}\right)+|\vec{d}|\\
r^T_N&=&r^S_N+\frac{\lambda^2}{\pi}\overline{b}^D d_{333}.
\label{eq:enocT}
      \end{eqnarray}
Those equations are valid for \linbo~ (point group $3m$), if the grating vector $\vec{K}$ is parallel to the trigonal $c$-axis and the vectors $\vec{\mu},\vec{E},\vec{k}$ are mutually perpendicular. Moreover, the approximation $e_{333}/C^E_{3333}\approx d_{333}$ was made. The sign in front of the dielectric constant in (\ref{eq:enocS}) is determined by the direction of the neutron spin $\vec{\mu}$: $+$ for parallel and $-$ for antiparallel to $\vec{E}(\vec{x})\times\vec{k}$. In another geometry with $\vec{\mu}\perp[\vec{E}(\vec{x})\times\vec{k}]$ this term even vanishes.
 The complete expressions are given by Equation (12) in Ref.  \cite{Havermeyer-apb99}. 
In comparison to the electrooptic coefficient for light, the situation is reversed in this case: $r^S_N\ll r^T_N$. This is again reasonable and reflects the corresponding contributions to the neutron-optical potential. We therefore suggest discontinuing the use of terms like ``primary'' or ``true'' for $\tens{ r}^S$ and ``secondary'' electrooptic coefficient for $\tens{ r}^T$, which can be found in literature.
An estimation of the various contributions to the electro neutron-optic coefficient for \linbo~ is summarised in Table \ref{tab:2}.
      \begin{table}\centering
\begin{tabular}{|l|c|}\hline
Contribution & [fm/V] \\ \hline
$r_N^T-r_N^S$& 3 \\
Schwinger-term &   $\pm 0.02$ \\
Foldy-term &  $-2\times 10^{-6}$ \\
EDM &  $<10^{-9}$ \\\hline 
\end{tabular}
\caption{\label{tab:2} Contributions to the ENOC for \linbo~ in  [fm/V] for cold neutrons with $\lambda=2$ nm and a grating spacing $\Lambda=400$ nm.}
      \end{table}

In analogy to light optics and  (\ref{eq:eo}), the neutronrefractive-index change induced by a holographically created space-charge field $E_{sc}$ amounts to
      \begin{equation}\label{eq:neo}
  \Delta n_N(x)=-\frac{1}{2} r_N E_{sc}(x).
      \end{equation}
In section \ref{sec:enoc} we will present the first experimental results on measurements of electro neutron-optic coefficients. 
\section{Experiments\label{sec:experimental}}
\subsection{Neutron experimental setup\label{sec:nsetup}}
\begin{figure}
  \centering
  \apbgraph{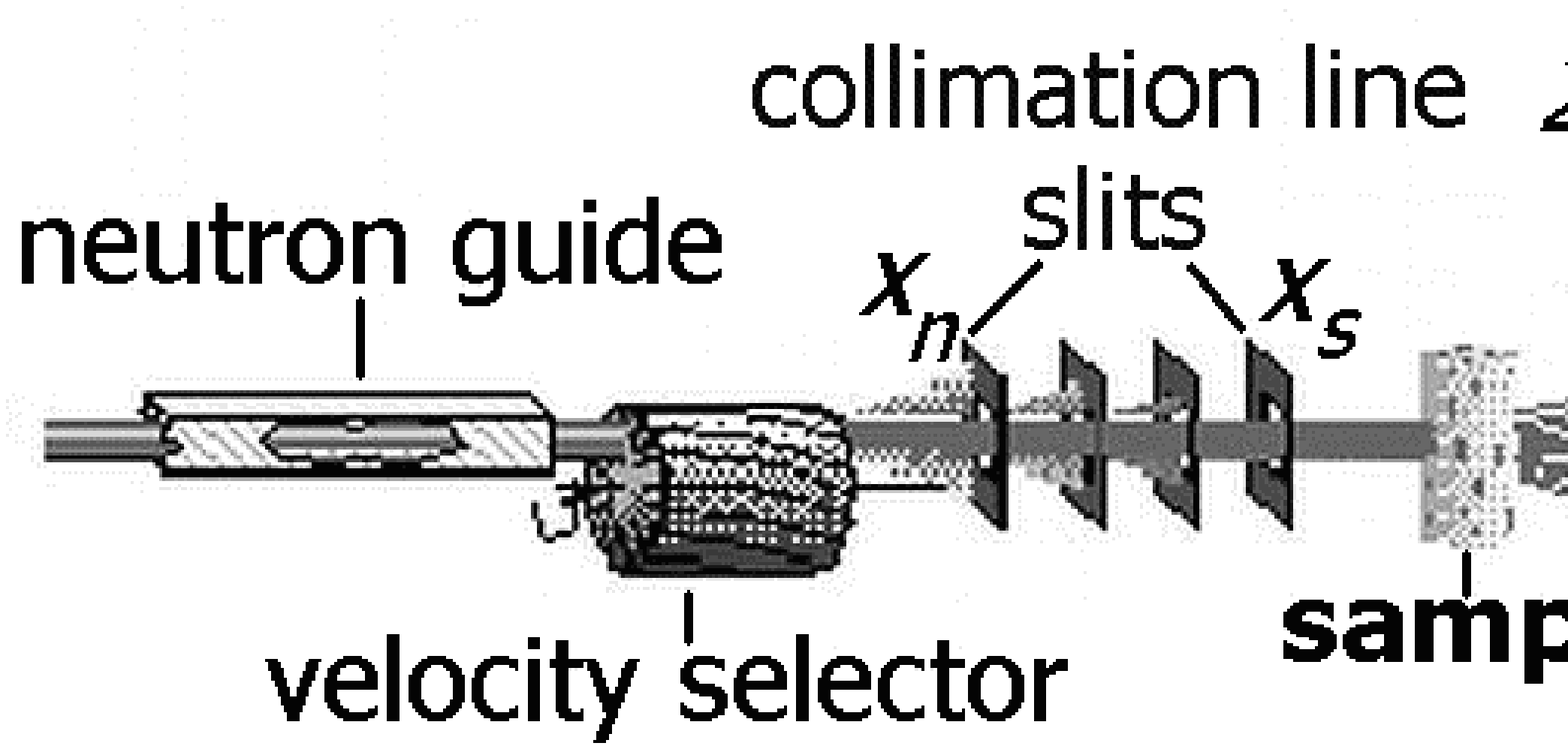}
  \caption{Measurement setup for neutron diffraction from light-induced neutronrefractive-index gratings. The sample is placed on a rotation stage. Typical collimation lengths $L_c$ are about 15 to 20 m. (Graphics  by courtesy of the Research Centre GKSS in Geesthacht, Germany)\label{fig:sans}}
      \end{figure}

\begin{figure}
  \centering
  \apbgraph{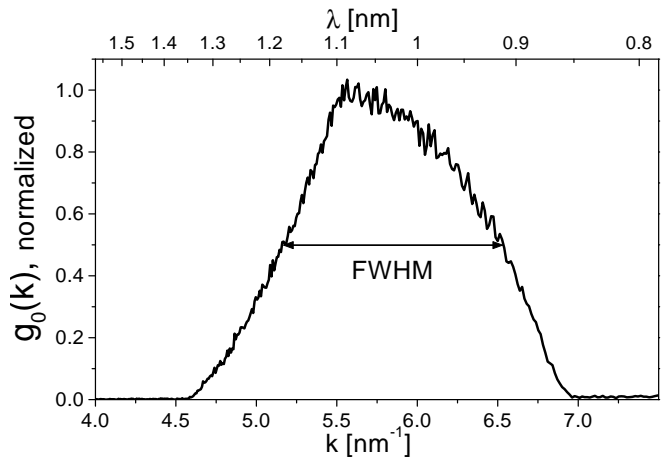}
  \caption{Normalised longitudinal momentum distribution $g_0(k)$ for cold neutrons with a central wavelength of $\lambda=1.1$ nm at the GeNF facility \cite{Havermeyer-00}. \label{fig:tof}}
\end{figure}

Because of the fact that the artificially produced grating constants are of the order of several hundred nanometers and that we are employing cold neutrons, the corresponding Bragg angles are a few tenths of a degree. Therefore, we utilise small-angle-neutron-scattering facilities (SANS). 
In a typical diffraction experiment at first the photo-induced neutronrefractive gratings are adjusted to obey the Bragg condition for neutrons. Then the diffracted and forward diffracted intensities are measured as a function of time and/or of the deviation from the Bragg-angle by rotating the sample. The diffracted and transmitted neutrons are monitored with the help of a two-dimensional position sensitive detector. The setup is depicted schematically in Figure  \ref{fig:sans}.
Aside from the neutron flux which is defined by the available neutron-source, two important experimental parameters are the collimation of the beam and the properties of the velocity selector. 
The first determines the spread of angles $\Delta\theta$ impinging on the sample, the second the wavelength distribution $\Delta\lambda$. 
The collimation can be tuned by using slits along the neutron beam path. We typically use a rectangular entrance-slit of width $x_n$ and a slit just in front of the sample $x_s$. To ensure a sufficiently collimated beam, the distance $L_c$ between the slits ranges from about 15 (SANS-2 at the Geesthacht Neutron Facility=GeNF) up  to 40 meters (D11 at the Institute Laue Langevin=ILL). The wavelength distribution can also be adjusted to the experimental needs. However, in practice, we optimise these parameters according to minimum demand (coherence properties) on one the hand and convenience (measuring time) on the other. Figure  \ref{fig:tof} illustrates the longitudinal momentum distribution $g_0(k)$ for cold neutrons at the GeNF facility.
The angular  (transverse momentum) distribution $g_{trans}(\Delta\theta)$ forms a trapezoid with a base $\Delta\theta_b=
|x_n+x_s|/L_c$ and a top $\Delta\theta_t=|x_n-x_s|/L_c$. Typical values are: $\Delta\lambda/\lambda\approx$10\% and $\Delta\theta=(\Delta\theta_b+\Delta\theta_t)/2<1$ mrad.
      \subsection{The early experiments (History)}
\begin{figure}
  \centering
  \apbgraph{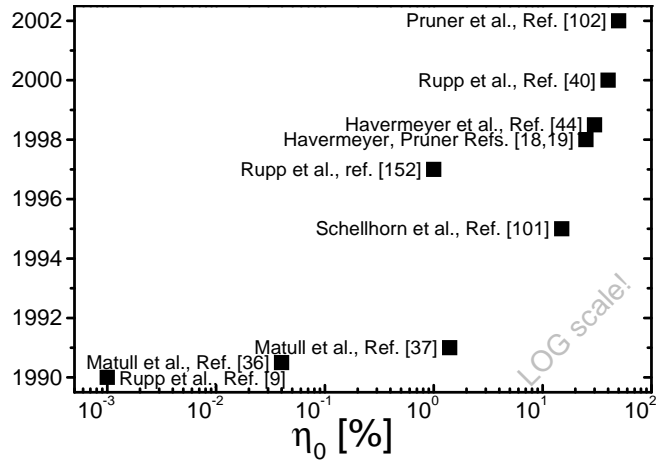}
  \caption{Progress in the production of artificial gratings for neutrons employing the photo-neutronrefractive effect. Maximum achieved diffraction efficiency for $\lambda\approx1.1$ nm \label{fig:progress}}
\end{figure}

For the first time a photo-neutronrefractive material was realised by Rupp et al. using a plate with a thickness of $\approx$ 2 mm consisting of PMMA. This PMMA-matrix was containing residual monomer and a photoinitiator which was sensitive in the visible wavelength region. The grating ($\Lambda=362$ nm) was prepared as described in section \ref{sec:holography}, using a recording wavelength $\lambda_p=514$ nm of an Argon-ion laser. The diffraction efficiency for $\lambda=1$ nm neutrons was in the range of $\eta_0=10^{-3}$\% \cite{Rupp-prl90}. Interpreting  (\ref{eq:nu}), (\ref{eq:densvar}) and (\ref{eq:nref}) it becomes evident how to improve the diffraction efficiency of the gratings:
\begin{equation}\label{eq:nuneut}
  \eta_0=\sin^2{(\nu)}=\sin^2{\left(\frac{\lambda d\overline{b}\Delta\rho}{2\cos{\Theta_B}}\right)}.
\end{equation}
In a follow-up publication the neutron wavelength, the sample thickness and the grating spacing were varied to reach maximum diffraction efficiencies of about 0.05\% \cite{Matull-zpb90}. The major limitation of those early attempts was the use of protonated PMMA. One of the reasons was the high incoherent scattering cross section of hydrogen so that the transmission was only around 20\%. The straightforward consequence suggested by the authors, namely using the deuterated analogue d-PMMA instead, was realised soon afterwards \cite{Matull-epl91}. In addition, a new uv-photosensitiser was introduced. The latter two have proved favourable for three reasons: incoherent scattering was reduced, the coherent scattering length density was increased (see Table \ref{tab:nref}), and the grating spacing $\Lambda$ could be easily reduced. As a consequence the maximum diffraction efficiency increased by a factor of 30 ($\eta_0\approx$1.5\%). However, the attempt to record a grating spacing with $\Lambda=120$ nm by employing counter-propagating recording beams  again led to extremely low diffraction efficiencies \cite{Matull-epl91}. At that time this fact was attributed to the restricted resolution of d-PMMA as a holographic recording medium. Rapid progress when trying to improve the quality of the photo-neutronrefractive samples currently allows diffraction efficiencies of up to 50\% for $\lambda=2$ nm neutrons (Figure  \ref{fig:progress}).
\subsection*{Applications}
These first successful experiments have opened up a wide field of potential applications:
\begin{enumerate}
\item By following the kinetics of relaxation processes in photopolymers, information on the complex phenomenon of glass-forming processes and polymerisation can be obtained (materials scientific aspect, see secs. \ref{sec:2harm}, \ref{sec:timeev}, \ref{sec:dyndiff}).
\item By diffracting the neutrons in the presence of an electric field, fundamental properties of the neutron itself are revealed (pure physics, see section \ref{sec:enoc}).
\item Utilising the knowledge about how to produce gratings for cold neutrons, neutron-optical devices can be designed. Mirrors, beam splitters, lenses or an interferometer are of outstanding technological relevance these days (technological aspect, see \ref{sec:holons}, \ref{sec:nif}). 

Those instruments then can again be used in turn to obtain material parameters (e.g. scattering lengths) or insight into the foundations of quantum physics (e.g. EDM).
\end{enumerate}
\subsection{Temporal evolution of the polymerisation\label{sec:timeev}} 
In order to obtain information about the kinetics of polymerisation it is important to systematically improve the production process of light-induced holographic gratings in d-PMMA (see. section \ref{sec:PMMA}). Though neutron scattering is a frequently applied method in polymer physics \cite{Higgins-94}, it has not yet been employed to study the temporal evolution of polymerisation processes over large time scales.

In the photopolymer system d-PMMA/DMDPE the recorded grating itself is used as a sensor to directly follow the glass forming process. Up to the last few years only diffraction experiments with light have been performed, which has turned out to be a favourable technique because of its simplicity. The refractive index for light $n_L$ in an isotropic material is known as
\begin{equation}
  \frac{n_L^2-1}{n_L^2+2}=\frac{4\pi N_A}{3}\overline{\rho}\,\overline{\Pi},
\end{equation}
according to the Lorentz-Lorenz relation, 
with Avogadro's constant $N_A$ and the mean polarisability $\overline{\Pi}$. Small photo-induced changes of $n_L$ therefore may be result from changes either in the density $\Delta\rho$ or the polarisability $\Delta\Pi$:
\begin{equation}\label{eq:dnlight}
  \Delta n_L=\frac{2\pi N_A (n_L^2+2)^2}{9n_L}\left(\Pi\Delta\rho+\rho\Delta\Pi\right).
\end{equation}
When the first diffraction experiments from light-induced gratings in PMMA were performed successfully, a technique became available which made it possible to receive complementary information to light diffraction. Combining  (\ref{eq:densvar}) and (\ref{eq:nref}) yields
\begin{equation}\label{eq:dnneutron}
\Delta n_N=\frac{\overline{b}\lambda^2}{2\pi}\Delta\rho.
\end{equation}
Thus, neutron diffraction is complementary in the sense that according to  (\ref{eq:dnneutron}) only density changes $\Delta\rho$ are probed, whereas light yields information on the combination of density and polarisability changes, cf.  (\ref{eq:dnlight}). Therefore light and neutron diffraction from photo-induced refractive-index changes can serve as a tool to clarify several aspects of the polymerisation in PMMA. 

To study the kinetics of the polymerisation process, light and neutron diffraction experiments were performed. 
As the glass forming processes are  irreversible, soon the demand for a facility was addressed which allowed for the simultaneous performance of light- and neutron-optic experiments. This led to the development and the design of HOLONS, which will be introduced in section \ref{sec:holons} in detail.

\begin{figure}
\begin{center}
  \apbgraph{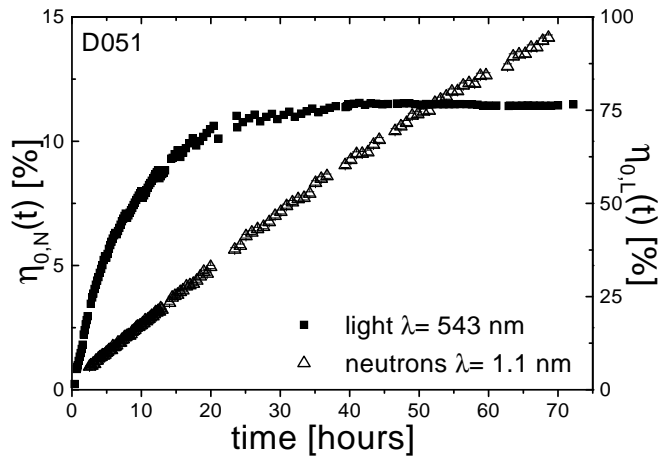}
  \caption{\label{fig:timeevn} Time dependence of the diffraction efficiency for neutrons $\eta_{0,N}(t)$ (left scale) and light $\eta_{0,L}(t)$ (right scale) at the exact Bragg-angle respectively. The measurement was performed after re-starting the polymerisation by illumination with uv-light for 10 seconds \cite{Havermeyer-00}. The reading wavelengths for neutrons were $\lambda=$ 1.1 nm and $\lambda=543$ nm for light respectively. Sample: d-PMMA D051}
\end{center}
\end{figure}

\begin{figure}
  \centering
  \apbgraph{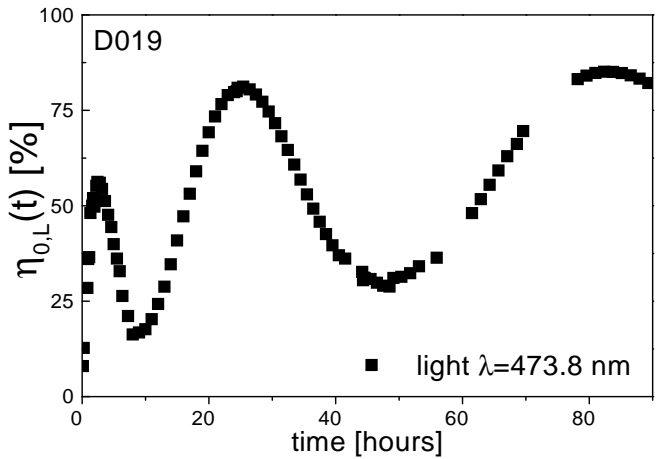}
  \caption{Diffraction efficiency $\eta_{0,L}(t)$ for light at the exact Bragg-angle for a sample with better homogeneity as compared to sample D051 shown in Figure  \ref{fig:timeevn} \cite{Havermeyer-00}. $\lambda=473.8$ nm. Sample: d-PMMA D019. \label{fig:D019}}
\end{figure}

At first measurements of the diffraction efficiency at the Bragg-angle $\eta(\Theta_B,t)=\eta_{0,N}(t)$ as a function of time are presented. 
Starting the observation immediately after the illumination of the photosensitive sample during the first hours, we were not able to detect any diffraction signal as the density changes according to  (\ref{eq:dnneutron}) develop rather slowly. In addition, the diffraction efficiency for light $\eta_{0,L}(t)$ was recorded as a function of time.
 Figure  (\ref{fig:timeevn}) shows the temporal evolution of $\eta_{0,N}(t)$ and $\eta_{0,L}(t)$ for the first three days after photo-polymerisation had started.
Using  (\ref{eq:diffractionefficiency}), now one could in principle easily determine the refractive-index changes $\Delta n(t)$, which reflect the kinetics of the polymerisation process. However, it is evident that the curves shown in Figure  \ref{fig:timeevn} do not at all obey a $\sin^2{(\nu)}$-function. This can be explained by inhomogeneities along the sample thickness and across the sample area \cite{Danckert-94,Breer-95,Kuhnt-97} and to the wavelength distribution. Fluctuations in the refractive-index change or of the grating spacing lead to a decrease of the contrast between the maxima and minima of the rocking-curve. This has a huge influence at the exact Bragg condition, where the extrema in the curve are smeared out until complete disappearance. In Figure  \ref{fig:D019} the diffraction efficiency is shown as a function of time for another d-PMMA sample (D019) with a better homogeneity. The curve comes here much closer to the sinusoidal behaviour we had expected.
But the major aim of the experiment actually is to study the kinetics of the polymerisation process by monitoring the refractive-index change $\Delta n$. Up to now we have only presented the diffraction efficiency at the Bragg-angle. Of course we realise that this is convenient. However, even for light $\eta_0$ does not allow to determine the refractive-index change unambiguously because of the periodicity in $\nu$ and the influence of the inhomogeneity of the refractive-index grating. In neutron diffraction the situation is even more complex as $\eta_{0,N}$ additionally depends significantly on collimation and wavelength distribution. The fact that the measurements are conducted with a partially coherent neutron beam, however, calls for the complete rocking-curve to extract the refractive-index changes unambiguously. Moreover, time resolved experiments and simultaneous light diffraction would till need to be achieved. Therefore the strategy is to measure rocking-curves $\eta(\theta)$ from time to time (1h for light, 10h for neutrons) to ensure the correct absolute value of $\Delta n$ and to monitor $\eta_0$ in the meantime. Further, a relation between the integrated diffraction efficiency $I(\nu)$ of  (\ref{eq:BWGint}) and $\eta_0$ is derived, which takes the influence of the partial coherence for neutrons into account.

\begin{figure}
  \centering
  \apbgraph{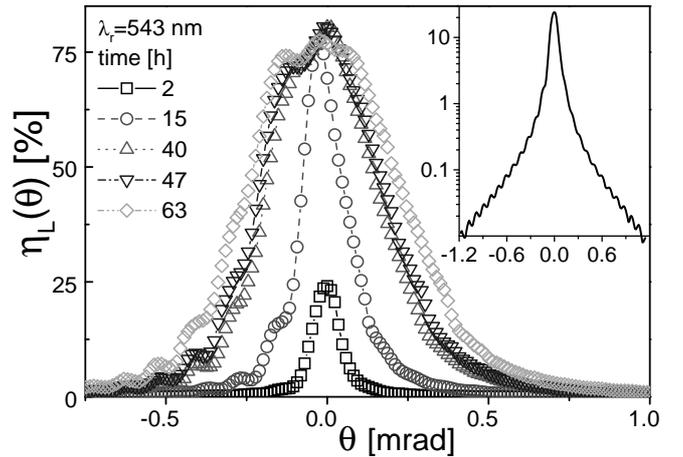}
  \caption{Rocking curve $\eta_L(\theta)$ for light at various times after starting the photo-polymerisation process \cite{Havermeyer-00}. The reading wavelength was $\lambda_r=543$ nm, the sample D051. The inset shows the angular dependence of the diffraction efficiency after $t=2$ h in a logarithmical scale. Note that thereby numerous side-maxima and minima become visible.\label{fig:rockL}}
\end{figure}

\begin{figure}
  \centering
  \apbgraph{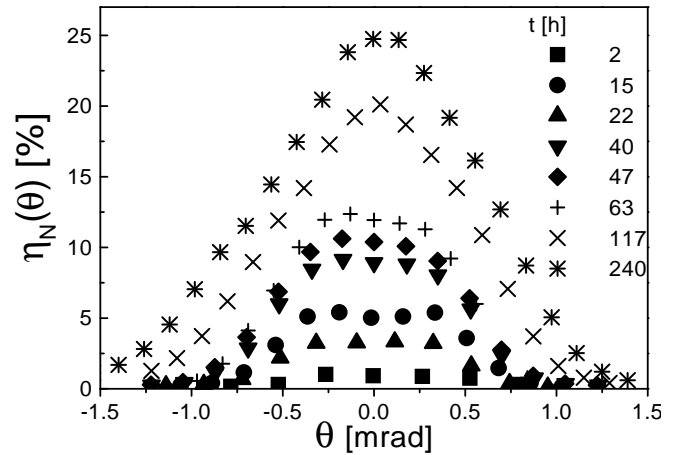}
  \caption{Rocking curve $\eta_N(\theta)$ for neutrons at several times after starting the photo-polymerisation process \cite{Havermeyer-00}. $\lambda=1.1$ nm,  $L_c$=14.4 m, $x_n$=15 mm, $x_s=$2 mm, $\Delta\lambda/\lambda=$0.23, sample: D051.  \label{fig:rockN}}
\end{figure}

Rocking curves for light are shown in Figure  \ref{fig:rockL} at certain times after initialising the refractive-index change.
The angular dependence of the diffraction efficiency does not simply follow  (\ref{eq:diffractionefficiency}), a fact which we have already learned from $\eta_{0,L}$. However, a faint periodic structure can be distinguished, in particular when inspecting the inset. Though the grating is obviously inhomogeneous, the effective thickness $d$ can be determined by calculating the difference between two neighbouring high order minima, as $\lim_{i\to\infty}\left[\theta_i^{(min)}-\theta_{i+1}^{(min)}\right]=\Lambda/d$. In general, the effective thickness which enters the equations turns out to be less than the geometric thickness. By performing a least-squares fit using  (\ref{eq:diffractionefficiency}) and (\ref{eq:BWGint}), the refractive-index change $\Delta n_L(t)$ remains the only free parameter. In Figure  \ref{fig:rockN} the corresponding measurement for neutrons is shown.
When interpreting Figure  \ref{fig:rockN}, it is striking that the shape of the rocking-curve changes significantly during the exposure time of the gratings \cite{Havermeyer-phb00}. Starting with a trapezoidal shape which resembles the angular distribution, we finally end up with a triangular or Lorentzian shaped curve. 

\begin{figure}
  \centering
  \apbgraph{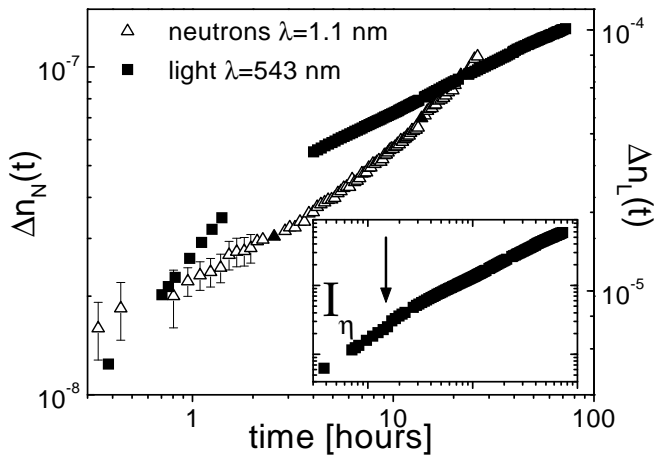}
  \caption{Time dependence of the refractive-index change for light $\Delta n_L(t)$(right scale) and neutrons $\Delta n_N(t)$(left scale) respectively. The solid triangles are determined from rocking-curves, the open symbols are calculated from measurements at the Bragg angle \cite{Havermeyer-00}. Sample: D051. $\Delta n_L$ was determined within the limits $\nu\ll \pi$ and $\nu\gg \pi$ (see text) using $I_\eta$, which is presented in the inset.  \label{fig:dnl+n}}
\end{figure}

At this point it seems necessary to account for the partial coherence of the neutrons, which is evidently the reason why the rocking-curves do not look like a $\rm sinc^2$-function. Assuming a normalised angular and wavelength distribution $g_0(k)$ and $g_{trans}(\Delta\theta)$ as discussed in section \ref{sec:nsetup}, the measured rocking-curve is the convolution with $\eta(\nu,\theta)$ from  (\ref{eq:diffractionefficiency}). The integrated diffraction efficiency 
\begin{equation}
I_{pc}(\nu)=\int g_0(k)\int g_{trans}(\Delta\theta)\left[\int\eta(\nu,\theta)d\theta\right] 
d\Delta\theta dk, 
\end{equation}
however, is independent on the angular spread. This is important as the integrated diffraction efficiency is accessible experimentally by integrating the rocking-curve measured. By $I_{\eta}$ we will denote the experimentally determined value multiplied by $d/\Lambda$ according to  (\ref{eq:BWGint}). Estimating the influence of the wavelength distribution up to second order in $\nu\Delta\lambda/\lambda$ yields \cite{Havermeyer-00}
      \begin{equation}\label{eq:BWGintpc}
  I_{pc}(\nu)=I(\nu)+\frac{J_0(\nu)-\nu J_1(2\nu)}{6}\left(\nu \frac{\Delta\lambda}{\lambda}\right)^2,
      \end{equation}
with $J_1(s)$ being the first order Bessel function. For the experimental reasons already discussed, it is important to relate $\eta_0$ with $I_{pc}(\nu)$, which is not possible in general. In the limit of small $\nu$ as well as large $\nu$, approximations can be found \cite{Havermeyer-00}.
Solving  (\ref{eq:BWGintpc}) for $\nu$ numerically now allows us to arrive at the refractive-index change $\Delta n_N$ for neutrons. In Figure  \ref{fig:dnl+n} the results for light and neutrons are depicted.
The temporal evolution of the refractive-index changes follows power laws quite well for light and neutrons. Havermeyer claimed the exponent to be the same and close to 1/2 for various samples investigated \cite{Havermeyer-00}. However, it is not possible to obtain unambiguous solutions when solving  (\ref{eq:BWGint}) or  (\ref{eq:BWGintpc}) for $\nu$. In the case of light the problem lies in the fact that due to the inhomogeneity of the grating in sample D051, the rocking-curve cannot be fitted using  (\ref{eq:diffractionefficiency}). As previously addressed, for neutrons, in addition, the problem of partial coherence arises. 
At the very beginning of the process (about 7 hours for light and 20 hours for neutrons), as long as $\nu\ll 1$ the approximation $I(\nu)=\nu^2$ holds and $\Delta n$ thus can be calculated. In the long term limit, i.e. the limit of large $\nu$, the integrated intensity is approximated by $I(\nu)=\nu/2$, and again $\Delta n$ can be evaluated.
For neutrons a least-squares fit of the function $\Delta n_N\propto t^{p_N}$ to the data yields for the exponent $p_N=0.45\pm 0.01$. For light, however, the time evolution of $I_\eta$ and hence $\Delta n_L$ clearly shows a distinct kink in the double logarithmic plot, indicated by the arrow in Figure  \ref{fig:dnl+n}, which separates regimes with different exponents of the power law. A least squares fit gives $p_{L1}=0.53\pm 0.004$ for $t<2$ h and $p_{L2}=0.766\pm0.001$ for $t>2$ h. In fact it is quite astonishing that several hours after an illumination with uv-light, which lasts for a few seconds only, the refractive index evolves over time spans of days! The first hours are then governed by an approximate $\sqrt{t}$-dependence of the changes for neutrons and light.
Further improvement in the sample production (at least of the quality of sample D019, see Figure  \ref{fig:D019}) must be achieved to either verify or falsify Havermeyer's assumption. 
Investigation of samples with different grating spacings $\Lambda$ lead to similar values for the exponents. This serves as a hint that diffusion does not play a decisive role in the photo-polymerisation process.

\subsection{HOLONS\label{sec:holons}} 
\begin{figure}
  \centering
  \apbgraph{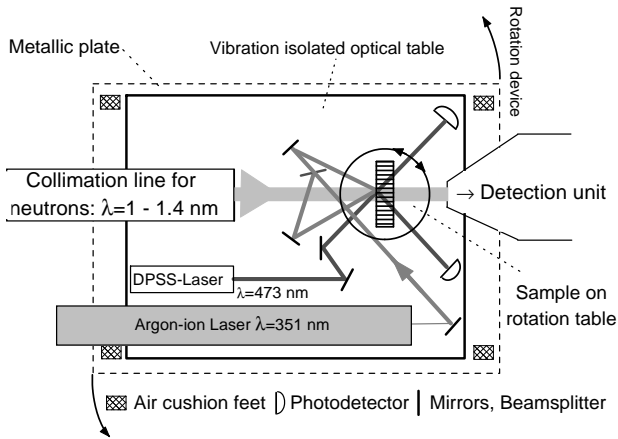}
  \caption{Sketch of the HOLONS experiment at the SANS-2. The light beams for recording the grating ($\lambda=$ 351 nm), for reconstructing ($\lambda=$ 473 nm) and the neutron beam ($\lambda=$ 1.1 nm) are indicated by grey lines. 
  \label{fig:HOLONS}}
\end{figure}

\begin{figure}
  \centering
  \apbgraph{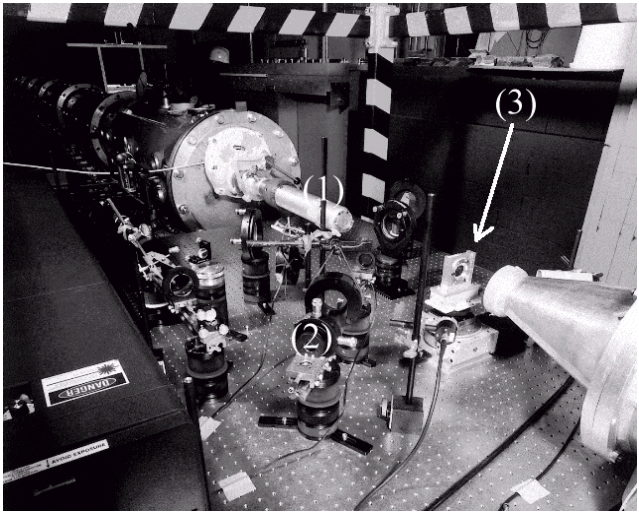}
  \caption{Picture of the HOLONS experiment. To the left the Argon-ion laser for recording the gratings and to the lower right the evacuated tube for the detection system can be seen. (1) marks the end of the neutron collimation system, (2) the beam expansion for the recording beams, (3) the photo-neutronrefractive sample (d-PMMA). Photograph by courtesy of  GKSS.\label{fig:holons-pic}}
\end{figure}

The measurements presented in the last section were performed using a novel experimental facility at the GeNF which was designed for {\bf time resolved simultaneous diffraction experiments with light and neutrons}. In addition it allowed us to utilise a complete holographic setup during conducting this type of experiments. The acronym HOLONS stands for {\bf Hol}ography and {\bf N}eutron {\bf S}cattering \cite{Rupp-spie98}. It basically consists of a holographic optical setup including a vibration isolated optical table, an Argon-ion laser ($\lambda=351$ nm) for recording the gratings and a diode pumped solid state laser ($\lambda=473$ nm) for reading. In Figure  \ref{fig:HOLONS} a sketch of the facility is presented.
The optical bench itself is placed on a metallic plate which can be transferred from the HOLONS-cabin to the cold neutron beamline SANS-2 by means of a crane. The HOLONS-cabin itself is situated in the guide hall of the GeNF. When performing an experiment, a rotation of the whole optical bench (1.2 tons!) with respect to the incident neutron beam is made possible by using air cushion feet and translation stages. Accuracy amounts to about $\pm$0.01 deg over a range of $\pm$ 2 deg. This technique ensures that the sample can be positioned in the correct Bragg angle for neutrons but keeps the light optical setup unchanged. In other words, neutron rocking-curves $\eta_N(\theta)$ can be performed while simultaneously measuring the diffraction efficiency $\eta_0(t)$ for light! This exactly meets our demand for clarifying the kinetics of photorefraction in doped polymers.
The photo-neutronrefractive sample is fixed on another rotation stage with high accuracy ($\pm$0.001 deg) in the common centre of both rotation devices. 
Because of the small Bragg angle for cold neutrons (1/10 deg), the neutron beam impinges nearly perpendicularly onto the sample surface. Therefore, the holographic recording geometry must be chosen in asymmetric configuration so that the beam splitter does not block the neutron beam. Thus the sample surface-normal is inclined with respect to the axis beam splitter-sample but still remains the bisector of the recording beams. Figure  \ref{fig:holons-pic} is a photograph of the HOLONS experiment at the SANS-2.
Summarising the benefits of HOLONS, we would like to emphasise its importance for
\begin{itemize}
\item improving and controlling the quality of photo-\-neutron\-re\-frac\-tive gratings
\item producing diffraction elements for calibration standards, beam splitters, lenses and further  neutron-optical devices on the basis of photo-neutronrefractive materials
\item performing time resolved simultaneous measurements of the diffraction efficiencies for light and neutrons
\item simultaneously recording holographic gratings and reconstructing them by light and neutrons.
\end{itemize} 
\subsection{Dynamic diffraction\label{sec:dyndiff}}
The improvement of the preparation technique led to samples which have diffraction efficiencies for neutrons of up to 50\% at the Bragg angle and thus $\nu\gg\pi$. Therefore, dynamical diffraction theory has to be applied and might be probed. For neutrons the decisive parameter $\nu$ can be extracted using  (\ref{eq:nuneut})
      \begin{equation}
\nu=\frac{\lambda d\,\overline{b}\Delta\rho}{2\cos{\Theta_B}}.
      \end{equation}
The validity of dynamical diffraction theory for thermal neutrons was demonstrated by changing the thickness $d$ \cite{Sippel-pl65,Shull-prl68} and the wavelength $\lambda$ \cite{Shull-prl68}. Though this has been a well-established subject for many years, in our experiments with photo-neutronrefractive samples $\Delta\rho$ and hence $\Delta n_N$ have been varied during the temporal evolution of the grating. Thus $\nu(t)$ has been continuously changed. Therefore, we expected the system to pass from the kinematical regime to the dynamical regime. However, because of the inhomogeneity of the samples and the broad wavelength distribution the typical Pendellösung oscillations were blurred out (cf. inset of Figure \ref{fig:etaint}). Moreover, the refractive-index changes for cold neutrons with wavelength $\lambda=1.1$ nm (Figures \ref{fig:rockN} and \ref{fig:dnl+n}) were too small to show a noticeable effect. To prove that actually dynamical diffraction theory must applied, wavelength dependent measurements similar to Shull's experiment \cite{Shull-prl68} were conducted.
\begin{figure}
  \centering
  \apbgraph{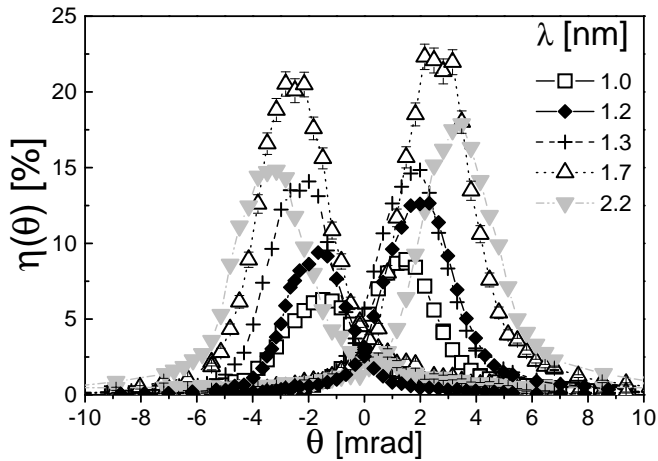}
  \caption{Probing dynamical diffraction theory: rocking-curves for different wavelengths 1 nm $\leq\lambda\leq$2.2 nm. The measurements were performed on sample D051 at the Paul Scherrer Institut (PSI) 20 months after production. \label{fig:lambda-psi}}
\end{figure}
After the evolution process of the gratings had approached a quasi-steady state, complete rocking-curves were collected for several wavelengths (Figure \ref{fig:lambda-psi}), thus allowing the calculation of the integrated diffraction efficiency $I_\eta$ (Figure \ref{fig:etaint}). 
Inspecting Figure \ref{fig:lambda-psi} we would like to draw the attention to the fact that the maximum diffraction efficiency $\eta_0$ at Bragg-angle does not occur for the wavelength $\lambda=2.2$ nm but for $\lambda=1.7$ nm. This clearly illustrates that kinematical theory fails to interpret the results correctly because $\eta_0(\lambda)$ is not a monotonous function of $\lambda$. In Figure \ref{fig:etaint} the even more significant measure $I_\eta(\lambda)$ is depicted for two different samples. Both samples qualitatively reveal the expected behaviour predicted by (\ref{eq:BWGintpc}). The integrated diffraction efficiency $I(\nu)$ deduced from the dynamical diffraction theory according to (\ref{eq:BWGintpc})  and (\ref{eq:BWGint}) are presented in the inset of Figure \ref{fig:etaint}. 
\begin{figure}
  \centering
  \apbgraph{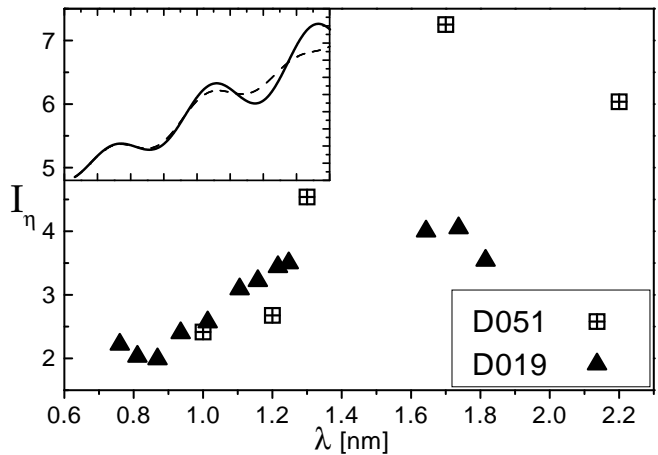}
  \caption{Integrated diffraction efficiency $I_\eta$ as a function  of the neutron wavelength for the samples D051 and D019 respectively. The inset shows the theoretical behaviour of $I(\nu)$ according to (\ref{eq:BWGint}) (solid line) and $I_{pc}(\nu)$ from (\ref{eq:BWGintpc}) (dashed line) for an assumed wavelength distribution $\Delta\lambda/\lambda=0.23$. \label{fig:etaint}}
\end{figure}
Unfortunately, the number of data points up to now has not been sufficient to unambiguously extract $\nu$ and hence the neutronrefractive-index changes. For the characterisation of the gratings as neutron-optical elements (beam splitter, mirror, interferometer) it is of paramount importance to obtain accurate values of the neutronrefractive-index changes. Supposing that sufficient time at a beamline will be at disposal, wavelength dependent measurements of $I_\eta$ may become the standard technique of characterising the neutron-optical elements based on the photo-neutronrefractive effect. In particular, the effective coherent scattering length density $\overline{b}\Delta\rho$ can be determined which is one of the figures of merit for neutron-optical devices. 
\subsection{Higher harmonics\label{sec:2harm}}
\begin{figure}
  \centering
  \apbgraph{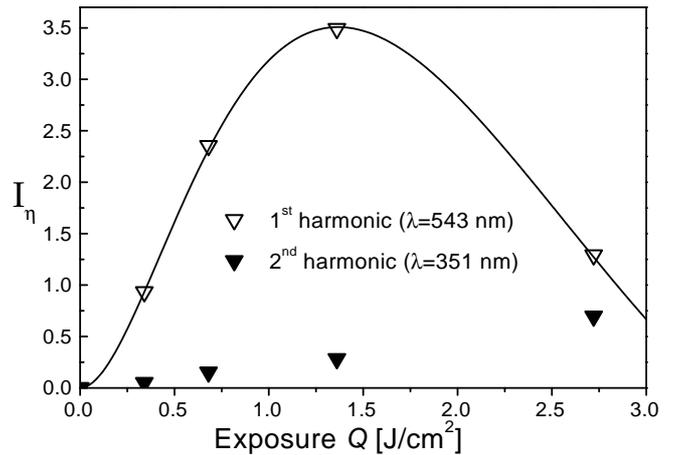}
  \caption{$I_\eta(Q)$ for a PMMA sample ($\Lambda=380$ nm) exposed to laser irradiation with $\lambda=351$ nm for different time spans (2 sec, 4 sec, 8 sec, 16 sec) and an intensity of $I_0=1700$ W/m$^2$. Rocking-curves were measured for the  first order Bragg peak (open symbol) and the second order Bragg peak(solid triangle) using light of wavelength $\lambda=543$ nm and $\lambda=351$ nm respectively \cite{Pruner-tbp02}. The solid line represents the first coefficient of $I(\nu)$ expanded to a Fourier series employing the proposed absorption model. \label{fig:abs-exp}}
\end{figure}

\begin{figure}
  \centering
  \apbgraph{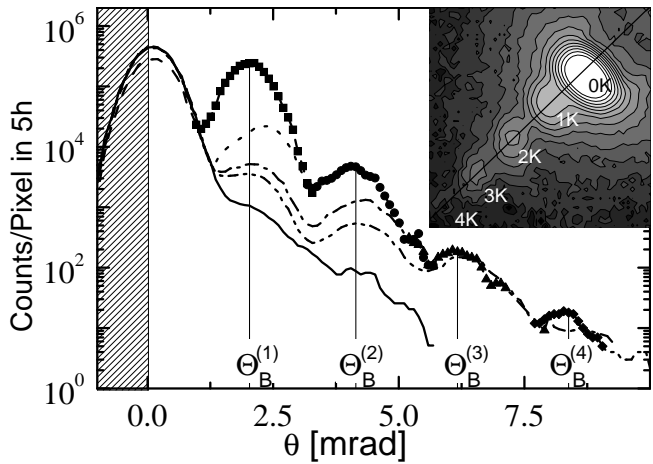}
  \caption{Counting rate along a horizontal line (=diagonal line in the inset) of the detector matrix as a function of the rotation angle for sample D019. Bragg peaks up to the fourth order are clearly visible. The full line represents an Off-Bragg measurement \cite{Havermeyer-00}. The inset shows part of the detector matrix with $0K, 1K, 2K, 3K, 4K$ denoting the corresponding diffraction orders. The sample was adjusted in the Bragg position for the fourth order (4K) \cite{Pruner-98}.\label{fig:high-harm}}
\end{figure}

Among the noteworthy characteristics of the photo-poly\-meri\-sation is its pronounced non-linear response to light exposure. This can be attributed to the growth and termination of the polymer chains in d-PMMA. Illuminating the photosensitised sample with a sinusoidal light pattern results in a refractive-index change which can be expanded in a Fourier-series according to  (\ref{eq:dn-harm}) with non vanishing higher Fourier coefficients. Performing diffraction experiments, this means that in addition to the (+1., -1.) diffraction orders higher harmonics appear. When the photorefractive effect in PMMA was studied by light, it turned out that the diffraction efficiency of the harmonics can be tuned by the proper choice of exposure. According to the model of the photorefractive effect in doped PMMA (section \ref{sec:absmodel}, \cite{Havermeyer-apb01}), the absorption changes and thus the refractive-index changes depend exponentially on the exposure $Q$. Only for very low exposure  can the response be approximated to be linear.  For comparison of the measured data with the results of the absorption model we used reasonable and appropriate parameters\cite{Havermeyer-apb01} for the rate and proportionality constants in (\ref{eq:abs-model}). Further we assumed that $\Delta n\propto\Delta\alpha$ and $I(x)=I_0 (1+\cos{(Kx)})$, and expanded $I(\nu(Q))$ into a Fourier series. The exposure dependence of the first and second Fourier coefficients then were compared to the measured values $I_\eta(Q)$. The experimentally obtained data and the results of the model are presented in Figure  \ref{fig:abs-exp}.
The constraint to obey the Bragg condition limits the possibility to observe diffraction from higher harmonics to $2 \Lambda^{(i)}>\lambda$, where $i$ denotes the diffraction order. Therefore, it was only possible to detect the second harmonic with light of $\lambda=351$ nm. This is inconvenient and may even damage the refractive-index profile of the sample as it is photosensitive in the uv-region. An elegant way of escaping this problem is to employ neutron diffraction instead. After d-PMMA samples with high diffraction efficiencies for neutrons had become available, Havermeyer et al. detected the second harmonic for three samples with grating spacings $\Lambda=400, 250, 204$ nm \cite{Havermeyer-prl98}. In the latter two samples the second diffraction order cannot be detected on principle by means of light. Nowadays higher harmonics up to the 4$^{th}$ diffraction order were observed, with a spacing $\Lambda^{(4)}=135$ nm \cite{Pruner-98,Havermeyer-00}. Figure  \ref{fig:high-harm} shows the counting rate as a function of the rotation angle along a horizontal line of the detector matrix. 
Moreover, the kinetics of the second harmonic was studied in detail similar to the procedure described in section \ref{sec:timeev}. There is evidence that the ratio of the density amplitude changes $\Delta\rho^{(1)}/\Delta\rho^{(2)}$ approaches a constant value after several hours. For the three samples investigated this value amounts from $(10 \ldots 25) \pm (2\ldots 10)$ \cite{Havermeyer-00}. These quite serious errors can be put down to the fact that for both harmonics complete rocking-curves had to be measured during their build up as well as to the fact that the second harmonic showed considerable diffraction efficiency only after several hours. This aspect still needs further investigation.

The controlled  setting of the magnitude of higher harmonics by tuning the exposure is important to facilitate the development of neutron-optical elements (see e.g. section \ref{sec:nif}). 
\subsection{The neutron interferometer\label{sec:nif}}
\subsubsection{Introduction and Motivation}
The extensive studies and experiments performed on the photo-neutronrefractive effect in d-PMMA and described in the previous sections served as a basis to set up a neutron interferometer utilising holographically produced gratings as beam splitters and mirrors. In our view the interferometer may be regarded as the most useful neutron-optical device as it provides information about the wave function, i.e. amplitude \textbf{and} phase, in contrast to standard scattering or diffraction techniques where only the intensity is measured. The successful development of a neutron interferometer \cite{Rauch-pla74} has hence led to a boost in neutron optics opening up completely new experimental possibilities in applied \cite{Schlenker-jmmm80,Rauch-nima87,Wallace-apl99} and fundamental physics \cite{Shull-prl80,Gahler-pra81,Rauch-hj81,Summhammer-pla82,Kaiser-prl83,Rauch-jdp84,Rauch-ps98}, e.g. extremely accurate measurements of scattering lengths \cite{Bauspiess-zp74,Boeuf-acb85,Kaiser-phbc86,Terburg-nima93,Ioffe-zpa94,Ioffe-pla97,Ioffe-pra98,Ioffe-prl99,Tomimitsu-pla00,Vrana-phb00} or tests of quantum mechanics \cite{Werner-PhT80,Greenberger-rmp83,Rauch-s93,Rauch-pb94}. For an excellent, complete and recent review on neutron interferometry see Ref. \cite{Rauch-00}. The neutron itself and its behaviour in various potentials, e.g. in a magnetic field \cite{Werner-prl75}, a gravitational field \cite{Colella-prl75,Greenberger-sa80,Horne-phbc88,Mashhoon-prl88,Werner-cqg94,Zouw-nima00,Zawisky-nima02} or such of pure topological nature  \cite{Cimmino-prl89,Allman-prl92,Allman-pra93,Wagh-prl97,Allman-pra97,Cimmino-nima00}, was studied as a model quantum mechanical system by means of the interferometer. Some very recent and amazing results on quantum states of the neutron in the gravitational field demonstrate the demand of further research on those topics \cite{Nesvizhevsky-n02}.
Coherence and de-coherence effects, which are important in any experiment, were investigated \cite{Kaiser-prl83,Klein-prl83,Rauch-pra96}. Based on the knowledge about the neutron and being interested in fundamental questions of physics, several groups started to elaborate interferometry with more complex quantum objects like atoms \cite{Ovchinnikov-prl99,Pfau-TD99,Inouye-n99,Mcguirk-prl00} and molecules \cite{Brezger-prl02}. On the other hand, the materials scientific aspect of neutron interferometry has not reached a satisfactory level up to now, i.e. it has not yet become established as a standard technique. Thermal neutron interferometers are run at the ILL (S18) and the NIST \cite{Jacobson-spie99}, which can be used for e.g. precise measurements of neutron scattering-lengths, but is naturally limited in wavelength by the use of perfect silicon crystals as beam splitters. Moreover, high expenditure (stabilisation against vibrations and thermal drift) was necessary to ensure phase stability and contrast \cite{ILL-yb01,Kroupa-nima00}. In the last few years investigations on biological materials and soft matter, e.g. complex organic molecules, membranes or bones, have met with more and more interest and the life science community has discovered neutron scattering as a non-destructive powerful method \cite{ILL-ar01,ILL-hr02}. Because of the large scale structures small angle scattering with cold neutrons is employed. This was one of the main reasons for the design of a new type of interferometer. The challenge was to build a low cost, flexible and versatile device for the operation at any SANS-facility, an interferometer for cold neutrons.
      \subsubsection{Design and methodology}
\begin{figure}
  \centering
\apbgraph{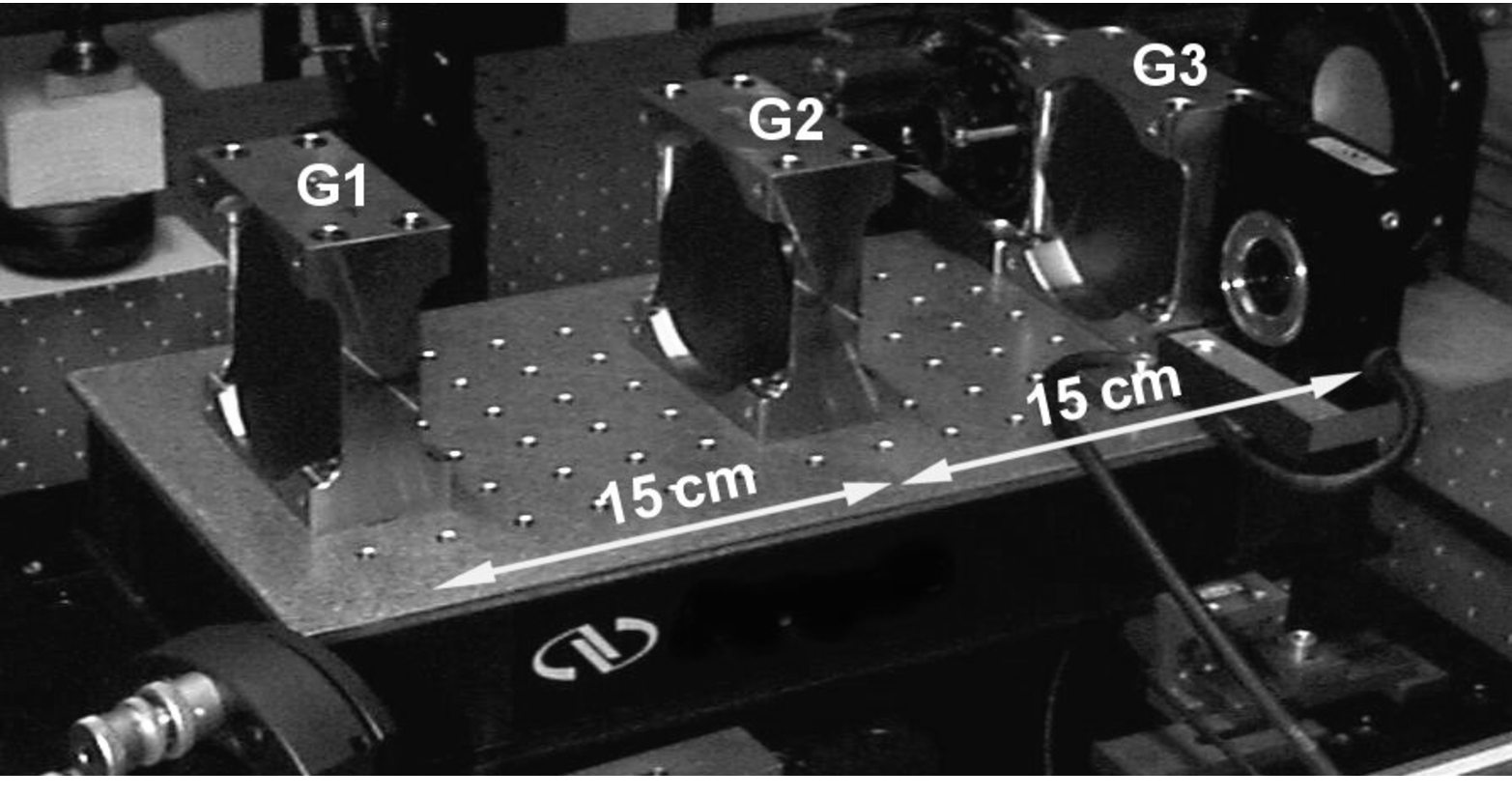}
\apbgraph{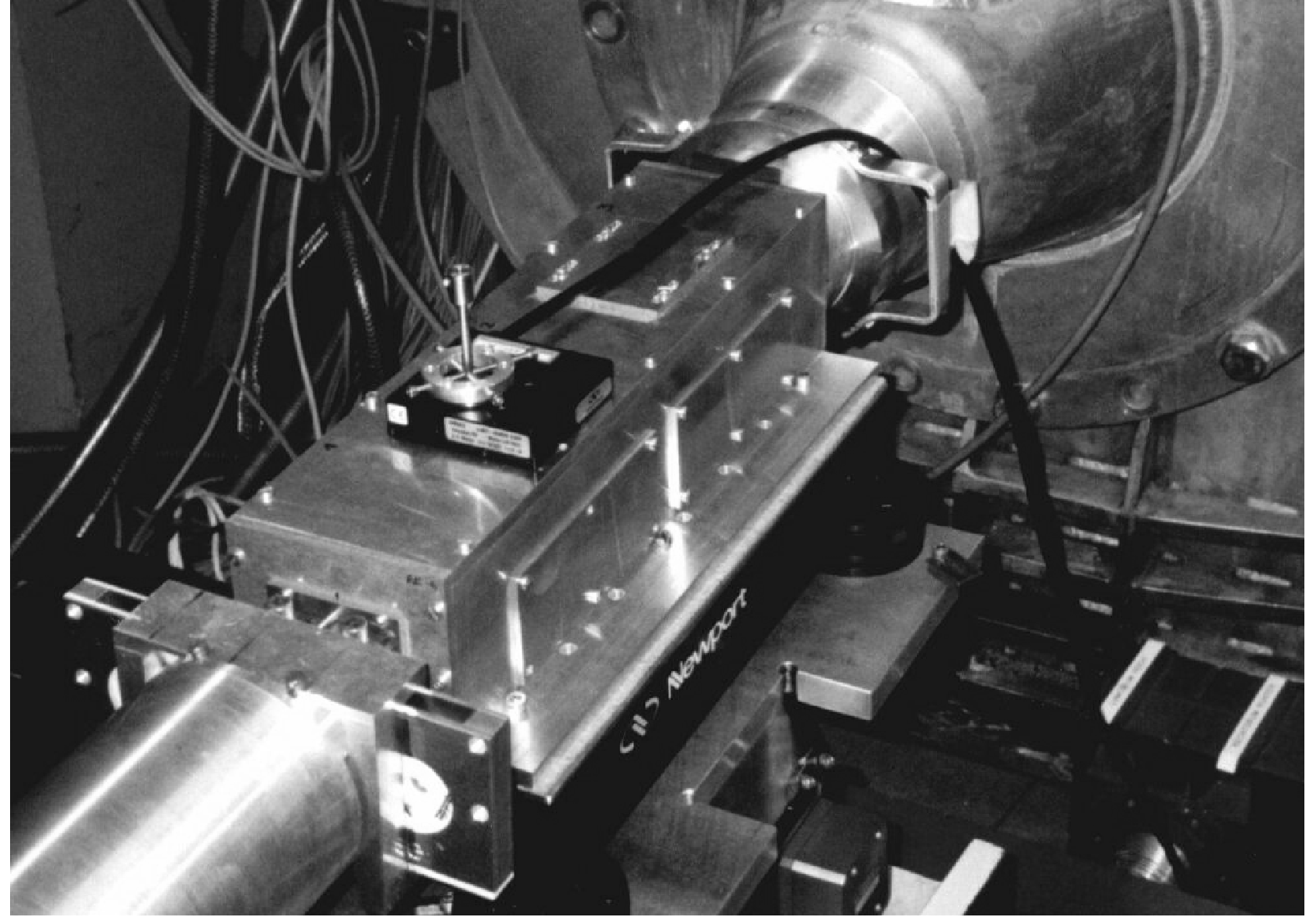}
\caption{The new interferometer for cold neutrons based on gratings in photo-neutronrefractive d-PMMA: after recording and aligning (top) and in the experimental environment of SANS-2 at GeNF (bottom). G1-G3 denote the gratings, L=150 mm, $\Lambda=380$ nm. \label{fig:nif}}
\end{figure}

\begin{figure}
  \centering
  \apbgraph{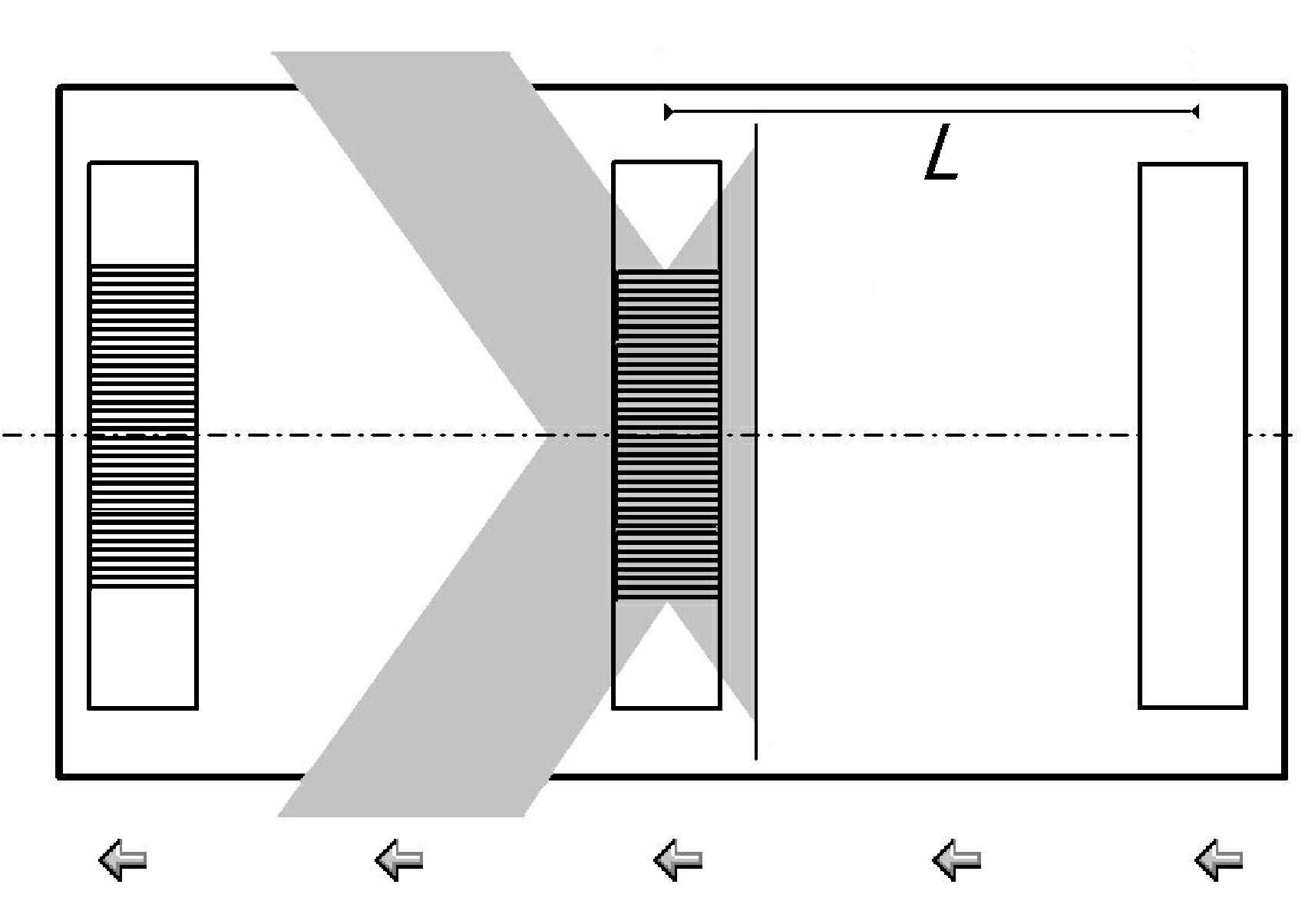}
    \apbgraph{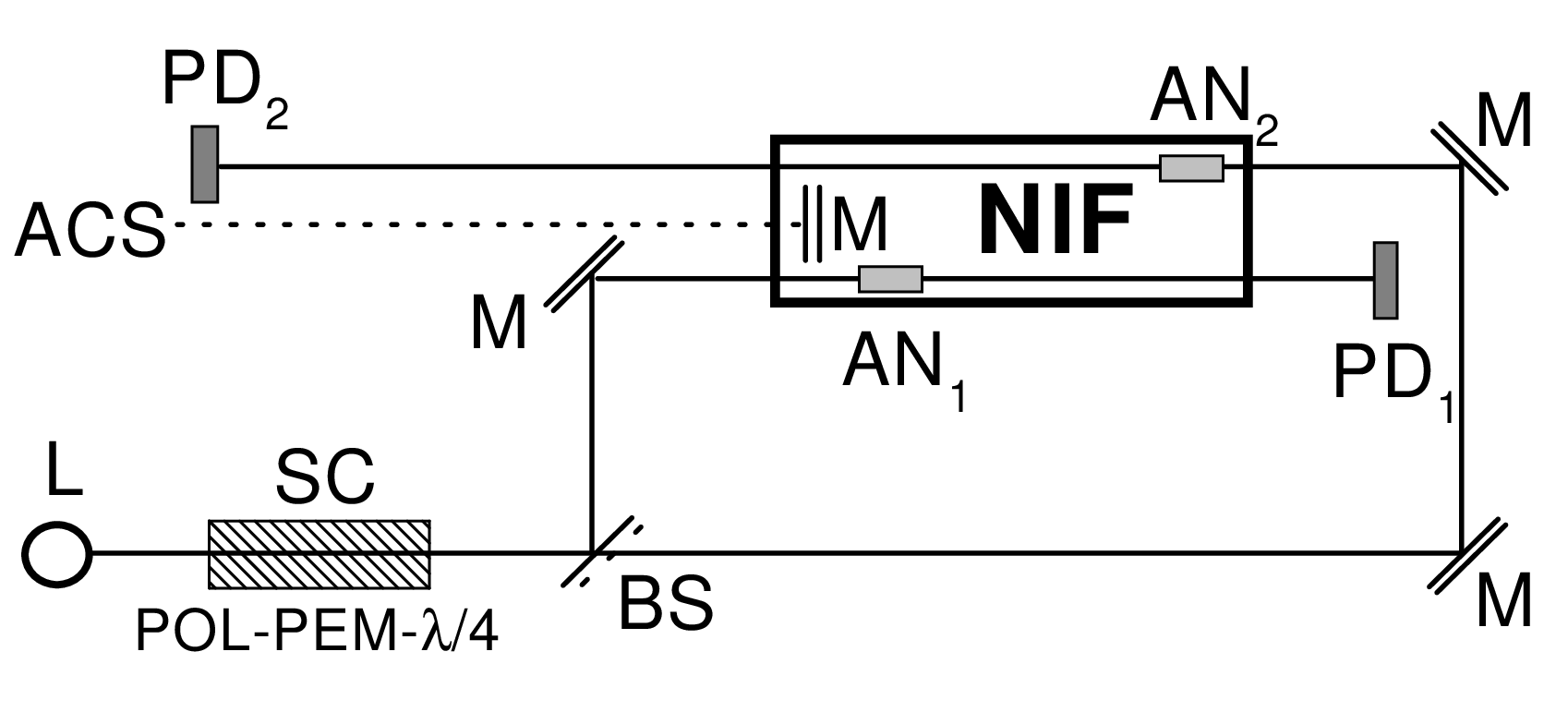}
  \caption{Sketch of the setup (top view) to produce an interferometer for cold neutrons. Illumination and translation of the photo-neutronrefractive samples to create gratings at distance $L$ (top). Light optic setup to control deviations from the ideal translation (bottom). ACS\ldots Auto collimation system (yaw, pitch) and polarisation optic setup to measure the roll angle: L\ldots laser, NIF\ldots neutron interferometer on translation stage, SC\ldots Sénarmont compensator consisting of a polariser, photoelastic modulator and quarter-wave plate, BS, M, AN, PD\ldots beam splitters, mirrors, analysers, and photodiodes, respectively. \label{fig:nif_setup}}
\end{figure}

The first neutron interferometer successfully run was designed by Rauch et al. \cite{Rauch-pla74}. This perfect-crystal neutron interferometer consists of three equally spaced parallel slabs that are produced by cutting two wide grooves in a large, perfect silicon crystal in the so-called LLL-geometry (triple Laue-case) \cite{Rauch-00}. Dynamical diffraction from the (220) reflection with a lattice constant of $\Lambda=0.19$ nm is used to split the incoming neutron beam and finally to recombine the sub-beams. Such an interferometer can be properly run with thermal neutrons, e.g. for wavelengths less than $\lambda\leq 0.6$ nm. Interferometers for very cold and ultra cold neutrons are based on different techniques: the gratings are created by sputter etching ($\Lambda\approx\mu$m range) in the LLL-geometry \cite{Summhammer-phb91,Gruber-pla89,Zouw-nima00}, by photolithography ($\Lambda\approx 20\mu$m) in reflection geometry \cite{Ioffe-jetpl81} or reflection from multi-layers is applied  \cite{Funahashi-pra96}. To close the gap between thermal neutrons and very cold neutrons Schellhorn et al. \cite{Schellhorn-phb97,Breer-95,Schellhorn-98} constructed a prototype interferometer in the LLL-geometry built of artificial gratings employing the photo-neutronrefractive effect of d-PMMA as described in the previous sections. They succeeded in demonstrating that the arrangement of the three gratings acts as an interferometer. Only recently was a new, larger and hence more sensitive interferometer constructed \cite{Pruner-tbp_nif02} and tested at the ILL and the GeNF. In Figure  \ref{fig:nif} a photograph of this interferometer is shown.
The crucial point for a successful operation of any LLL-interferometer is the extremely accurate mutual alignment of the gratings. Thus perfect silicon crystal interferometers rely on the quality of the silicon crystal. The cutting and etching of a monolithic crystal ensures the desired accuracy over the full distance. For the very cold neutron interferometer the alignment of the phase gratings is performed dynamically by tilting and translating each grating which is controlled by operating three auxiliary laser-light interferometers additionally \cite{Zouw-nima00}! Thermal and acoustic isolation for both types of interferometers is mandatory. The methodology to evade such problems and to construct an interferometer for cold neutrons is as follows: 
      \begin{itemize}
\item Three photosensitive d-PMMA samples are prepared and mounted on a linear translation stage. 
\item The first photo-neutronrefractive slab is exposed to the interference pattern of the holographic two-wave mixing setup (cf. Figure  \ref{fig:nif_setup}, top).
\item Successively the second and third slab are moved, i.e. nominally translated, to the position of the sinusoidal interference pattern. Prior to exposure the motion is corrected for deviations from the ideal translation (pitch, yaw and roll). The latter is the decisive step during the production of the interferometer as demand on accuracy is paramount. 
    \end{itemize}
Estimations of the loss in contrast due to errors in the alignment of the gratings are given in Refs.  \cite{Breer-95,Zouw-nima00,Weber-98,Zouw-00}. 
To control the accuracy of the translation and to correct deviations, an optical system was used (sketch in Figure  \ref{fig:nif_setup}, bottom) in combination with piezo driven stages. Details of this technique will be published elsewhere \cite{Pruner-tbp_roll02}. The roll angle is the most critical parameter. To control the latter, we developed a polarisation optic method which is independent of the distance of translation. Therefore in principle interferometers of any length may be fabricated using this technique.

The advantages of this attempt over other techniques are striking:
the grating spacing $\Lambda$ is easily tailored to the required value within the range 250 nm$<\Lambda<10 \mu$m. Moreover, the sinusoidally modulated light pattern creates - if properly prepared (see section \ref{sec:2harm}) - a sinusoidal grating. Therefore only  $+1.$ and $-1.$ diffraction orders occur. In addition, the adjustment of the three photopolymer samples is done during the recording of the grating once forever. Finally, the whole interferometer has turned out to be very stable, compact and robust despite our  high demand on accuracy. The prototype \cite{Schellhorn-phb97} was produced in Osnabrück (Germany), transferred to the neutron source of the ILL (Grenoble, France) and was set up ready to run within a few hours.
      \subsubsection{Characteristics, first results and outlook}
The characteristics of the recently constructed interferometer are summarised in table \ref{tab:nif}. The setup is symmetric in the LLL geometry with beam paths as illustrated in Figure  1 of Ref.  \cite{Schellhorn-phb97}. Due to small Bragg angles at cold neutron wavelengths the interfering 0-($=trr+rrt$) and H-($=rrr+trt$) beams cannot be separated easily from the parasitic beams ($=ttt, ttr, rrr, rtr$) and thus contrast and visibility $v$ are reduced. This obstacle can be overcome by producing larger interferometers with smaller grating constants, using longer neutron wavelengths or narrow slits. A further very promising possibility already under way is referred to at the end of the chapter.
Here, we denoted the amplitude of the transmitted and reflected plane waves at each grating as $t$ and $r$ respectively. The visibility $v$ is defined as \cite{Rauch-00}:
      \begin{equation}\label{eq:vis}
v=\frac{I_{Max}-I_{Min}}{I_{Max}+I_{Min}}=
m|\Gamma^{(1)}(\vec{x}-\vec{x}',t-t')|.
      \end{equation}
If we know the modulation $m$, direct experimental access to the absolute value of the normalised correlation function $\Gamma^{(1)}(\vec{x}-\vec{x}',t-t')$ and therefore to the coherence length $l_c$ is opened by measuring the interference fringes and their decay with increasing phase shifts.
\begin{table}
  \centering
\begin{tabular}{|p{\columnwidth/3+3mm}|l|rl|} \hline
Property & Symbol & Value & \\ \hline
Total length & 2$\times L$ & 300 & mm \\
Grating spacing & $\Lambda$ & 380& nm \\ 
Weight & & 15 & kg\\
Dimension & $l\times w \times h$ & $40\times 20\times 15$ & cm$^3$\\\hline
Enclosed beam area & $A_0\approx L^2\lambda/\Lambda$ & 1.2 &cm$^2$\\
Beam separation at third grating & $\approx L\lambda/\Lambda$& 0.8 & mm\\
Diffraction efficiencies & $\eta_{1,2,3}$, $\eta_{NIF}$ & $\approx$40\% & \\
Visibility & $v$ & 21\% &\\
Resolution & $\Delta E$ & $\leq 10^{-11}$ & eV \\    \hline
\end{tabular}
    \caption{Characteristic values for the new cold neutron interferometer (Figure  \ref{fig:nif}) for a wavelength of $\lambda=2$ nm.\label{tab:nif}}
\end{table}

\begin{figure}
  \centering
  \apbgraph{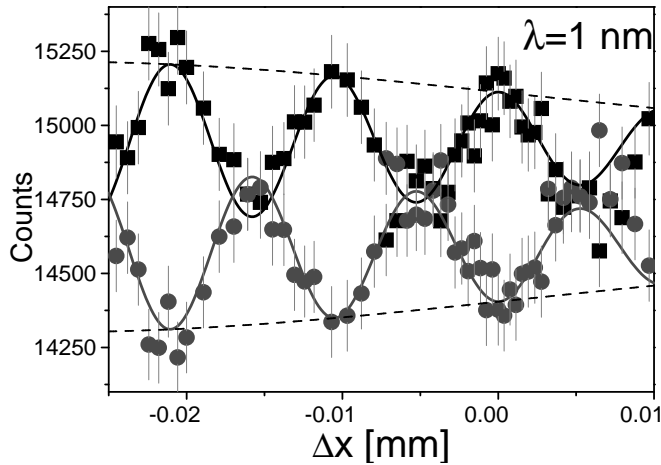}
  \caption{Part of an interferogram obtained by rotating a 4.1 mm thick phase flag of sapphire around an axis perpendicular to the plane of incidence. The measurement was performed at the GeNF with a neutron wavelength of $\lambda=1 $nm and a counting rate of about 6000/s. The solid lines are obtained from fitting (\ref{eq:interferogram}) to the data. The dashed lines represent the Gaussian envelope function which describes the decay of the visibility. \label{fig:interferogram}}
\end{figure}

The interferometer was operated one month after its production at the ILL (instrument D22) and 2 months later at GeNF (instrument SANS-2). The main purpose of those experimental terms was to test and characterise this large scale interferometer. The overall diffraction efficiency $\eta_{NIF}$ was measured by rocking the whole interferometer through the Bragg peaks. The $\pm 1.$ and $\pm 2.$ diffraction orders were visible with maximum diffraction efficiencies $\eta_{NIF}^{(+1)}=0.434$, $\eta_{NIF}^{(-1)}=0.354$, $\eta_{NIF}^{(+2)}=0.024$, $\eta_{NIF}^{(-2)}=0.021$ respectively for a wavelength of 1 nm. For the interferometric experiment the interferometer was adjusted to the Bragg angle with highest diffraction efficiency. Then a phase flag made from sapphire was rotated through both beam paths, thus yielding a relative phase shift between the 0- and the $H$-beam as previously described in Ref.  \cite{Schellhorn-phb97}. We expect an interference pattern of the form
      \begin{eqnarray}\label{eq:interferogram}  
I_{0,H}(\Delta x)&=&A_{0,H}+v\cos{\left(\overline{b}^D\lambda (\Delta x-x_0)\right)}\\
v&=&m\exp{\left[-\left(\frac{\Delta x-x_0}{\sigma}\right)^2\right]},\nonumber
      \end{eqnarray}
if a Gaussian coherence function is assumed. Here $\Delta x$ and $x_0$ are the geometrical path difference due to the rotation of the flag and the initial path difference of the empty interferometer respectively, $\sigma$ is the width of the wave packet and hence directly related to the coherence length $l_c$. The parameters $A_{0,H}$ and $m$ are functions of the diffraction efficiencies $\eta_{1,2,3}$ of the gratings. In addition, contrast is reduced because of inhomogeneities of the phase flag, thickness variations and beam attenuation effects \cite{Kaiser-prl83,Clothier-pra91}.
The resulting interference pattern is presented in Figure  \ref{fig:interferogram} for a wavelength of $\lambda=1$ nm (distribution similar to that shown in Figure  \ref{fig:tof}) and a collimation of $x_n=30$ mm, $x_s=1$ mm, $L_c=15.5$ m. A maximum visibility of 2\% is achieved. Figure  \ref{fig:interferogram} clearly demonstrates that the fringe visibility is continuously reduced upon larger phase shifts. This can be attributed to the diminished overlap of the partial wave packets traveling along two different beam paths. The decrease allows us to estimate the longitudinal coherence function according to  (\ref{eq:vis}). The (longitudinal) coherence length determined by the fit amounts to 10 nm.
Note that the initial phase of the interferometer is about $-5\pi$. We suppose that this be put down to a combination of the gravitational potential \cite{Staudenmann-pra80} (phase shift due to the Sagnac effect \cite{Werner-prl79} or the COW \cite{Colella-prl75}) and an internal phase because of inhomogeneities, imperfections and misalignment of the gratings. An initially asymmetric alignment of the sapphire flag might also be the reason. However, the latter is not very likely as it would correspond to a rotation angle of 55 degrees.
We also employed  larger wavelengths ($\lambda=2, 2.7$ nm) at a more powerful neutron source (ILL) and measured the longitudinal coherence function. The results will be published elsewhere \cite{Pruner-tbp_nif02}.

The present approach exhibits several inherent features that will open up new possibilities in neutron physics. The ultimate goal is to create a novel device for neutrons which can easily be implemented and run at any beam line for cold neutrons. The principal future application of such an instrument will be the investigation of mesoscopic structures and their kinetics in the fields of matter physics and engineering, chemistry, and biology. The latter is already today the field with the most frequent proposals for beam time at SANS installations. Seen the present status, it may sound revolutionary to add the interferometer to SANS instruments for routine phase measurements but the approach discussed above still allows for considerable improvement. So let us summarise its main advantages, discuss its future perspectives and its possible impact on physics and/or materials science:
\begin{itemize}
\item The spatial frequency $K$ of the interferometer is not fixed ab initio but can be adapted to the specific problem of the SANS investigation. In future it may even be possible to `record' the interferometer in real time, depending on the problem under investigation.
\item Polymer slabs permit us to record so-called volume holograms. Hence, several interferometer paths running at different $K$ values can be implemented at the same time into the same set-up. In addition, the superposition of gratings allow us to adapt the angular and spectral acceptance range of the white-beam interferometer, thus optimising the diffracted flux of the interfering beams.
\item A coherent small angle scatterer as the sample under investigation can replace the first grating of our LLL interferometer. We call this an SLL interferometer where S is the scatterer. Possible applications are the investigation of the slow kinetics of precipitation in metals, the kinetics of relaxation processes in glasses, or of light-induced processes in biological materials such as bacterio-rhodopsine by monitoring the changes of the neutron phase. Another application includes deciphering the structure of complicated biological samples by performing an additional phase analysis of the diffraction pattern. 
\item By shifting the analyser grating - this is the last one of the LLL gratings - small energy shifts can be detected. In that mode of an SLF interferometer (F=Fourier analyser grating) quasi-elastic SANS can be performed.
\item Performing standard interferometric experiments (see above) using a phase flag, the coherence properties of the cold neutrons can be probed. This is particularly important for scattering, diffraction, spin echo and reflection experiments \cite{Goldman-jap97,Keller-phb97,Sinha-prb98,Bernhoeft-prl98,Bernhoeft-aca99}, e.g. reflectometry with polarised neutrons where the coherence volume needs to be known and is a decisive quantity only roughly estimated \cite{Zabel-phb00,Kepa-prb01,Schreyer-pc,Donovan-prl02} or even more often simply not considered. 
\item As a consequence of the nonlinearities in the process of photo-polymerisation (see section \ref{sec:2harm}), gratings with periods as low as $\Lambda=100$ nm can be (and have in principle already been  \cite{Pruner-98,Havermeyer-00,Havermeyer-prl98}) achieved, i.e. such an interferometer spans the mesoscopic range from 100 nm to 10 $\mu$m.
\item Finally, we would like to recall that the LLL-geometry is in fact equivalent to that of a Mach-Zehnder interferometer, i.e. the first grating should serve as an ideal beam splitter (50:50), the second grating as a mirror (0:100), the third again as a beam splitter. In contrast to the attempts employing a perfect crystal interferometer from silicon, we are able to tune the properties of the corresponding grating during preparation. The diffraction efficiencies of the d-PMMA slabs depend on the thickness, the wavelength and the light-induced neutronrefractive-index change and can amount up to 80\%. Thus a nearly perfect situation may be reached by optimising the aforementioned parameters of the second slab, e.g. by doubling the thickness as compared to grating one and three. Another possibility would be the application of different illumination times (see section \ref{sec:2harm}). In the case that the second grating really acts as a mirror, the intensity of the parasitic beams will be reduced drastically, thus enhancing the visibility! Experiments for this improvement are already well under way.
    \end{itemize}
It is supposed that the flexibility, the low costs and the excellent properties of the interferometer composed of gratings created by the photo-neutronrefractive effect will promote the development of novel small angle scattering techniques with cold neutrons. Proof of the topicality of neutron interferometry, in particular with cold neutrons, is given by the fact that the planned \textit{European Spallation Source ESS} considers this type of experiments as flagship experiments \cite{Abele-ESS01}.
In addition, due to its different design and operating wavelength range, our type of interferometer can contribute constructively to unsolved problems, e.g. about the consistency of the measured gravitational phase shift with theory\cite{Bonse-prd84,Werner-phbc88,Littrell-pra97,Werner-jpsj96,Littrell-aca98,Zouw-nima00}. 
\subsection{Ageing of gratings in d-PMMA}
\begin{figure}
  \centering
  \apbgraph{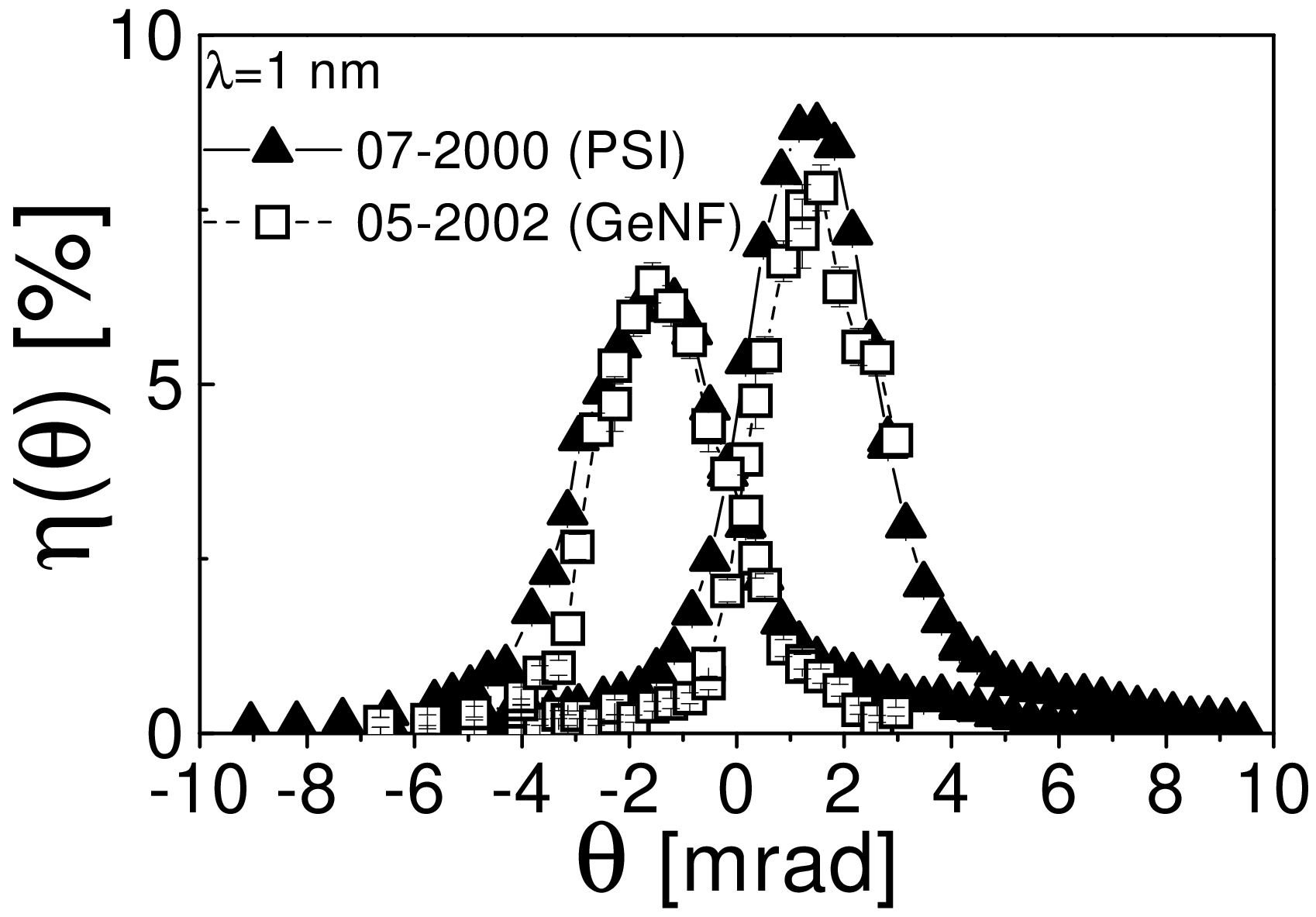}
  \apbgraph{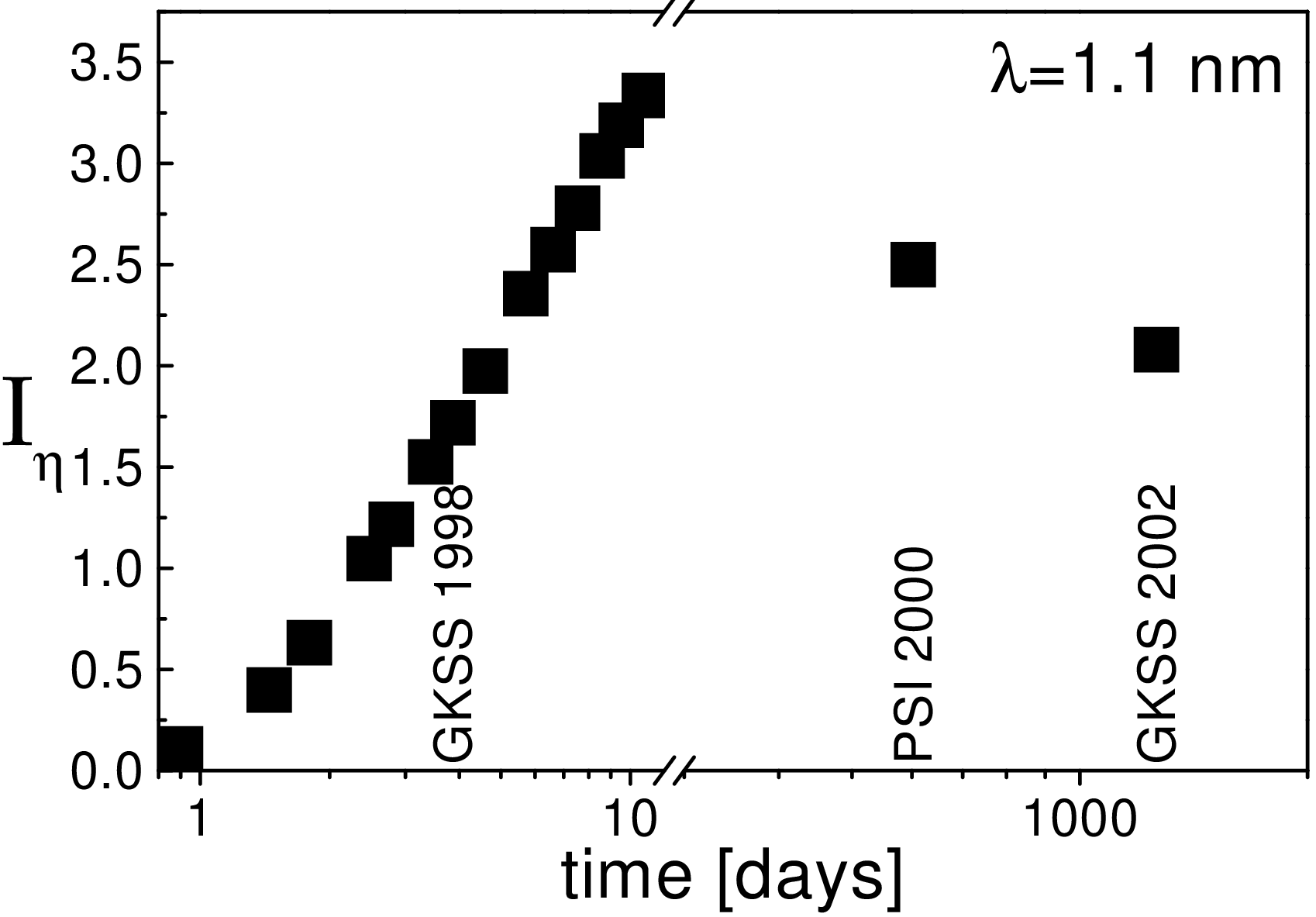}
  \caption{Ageing of the sample D051.\label{fig:aging}}
\end{figure}

Before introducing gratings in d-PMMA based on the photo-neutronrefractive effect as a standard neutron-optical component, the question of lifetime arises. Since the fabrication of gratings with diffraction efficiencies in the range of 10\%-50\% reached a satisfactory level only a few years ago, the experience has been rather limited. The oldest grating that is available and still can be used is sample D051 (recorded on 19-11-1998). This sample was kept in its chassis between the two glass plates, which turned out to be advantageous for its quality. The grating forming process was monitored for one week (see section \ref{sec:timeev}). In July 2000 and then in May 2002 rocking-curves were measured again. The results can be seen in Figure  \ref{fig:aging}: The diffraction efficiency increased for several weeks (followed also by light optical measurements) until it reached a certain limit ($I_\eta\approx 3.5-4.0$) and then decreased to a level of 60\% ($I_\eta\approx 2$) after 4 years. An additional problem might have been that the sample was inhomogeneous and thus the various measurements were not performed on the same area of the sample. We estimate this error to be 20\%, which corresponds to $\Delta I_\eta=0.4$ \cite{Pruner-98}.
Now (in July 2002), sample D051 became detached from the glass plates because of continuous  shrinkage due to polymerisation and/or temperature variation, i.e. the quality will decrease considerably. For future applications it will be favourable to automatically squeeze together the glass plates by a spring to avoid detaching. 
\section{Electro neutron-optics \label{sec:enoc}}
In this chapter we will present a standard electrooptic material (\linbo) exhibiting a photo-neutronrefractive effect of a completely different nature. The corresponding mechanism has already been briefly discussed in section \ref{sec:linbo} and is based on the creation of space charge fields which modulate the neutronrefractive index. Further, the experimental results obtained by diffracting neutrons from the light-induced gratings will be discussed. It was our primary aim to determine the ENOC of \linbo~ and to unravel  the contributions according to  (\ref{eq:enocS}) and (\ref{eq:enocT}) in further experiments.

Now one may ask why investigate the substances which are expected to show only small refractive-index changes (see table \ref{tab:2}, (\ref{eq:neo})) with diffraction efficiencies much less than in photopolymers. The reason is that the amazing perspectives opened up by the results of the corresponding experiments justify the performance of this difficult task: Measurements with polarised neutrons would allow us to detect Aharonov-Casher (spin-orbit) \cite{Aharonov-prl84} as well as Foldy contributions \cite{Foldy-pr51,Foldy-pr52} to the diffraction, and might even offer a novel experimental approach towards the fundamental question of the EDM of the neutron. Some of the effects have already been carefully studied  \cite{Cimmino-prl89,Allman-prl92,Allman-pra93,Cimmino-nima00,Alexandrov-92} using interferometric techniques. Moreover, during the last few years different experimental approaches were offered to lower the limit of a tentative EDM  \cite{Ioffe-nima84,Fedorov-phb97,Freedman-nima97,Lamoreaux2-nima99,Lamoreaux-nima99,Zeyen-nima00,Harris-nima00} including open debates, and this still remains a challenging topic.

This gives rise to the question why scientific literature does not refer to any measurement of ENOCs. The simple reason is that the neutronrefractive-index changes due to an applied electric field are rather tiny (cf. table \ref{tab:2}). Therefore, any standard technique (refraction, total reflection, interferometry, diffraction from a crystal lattice) must fail in detecting an electro neutron-optic effect. Recalling that the neutronrefractive-index in the thermal to cold neutron spectral range differs only by about $10^{-4}$ or less from the refractive index for vacuum, we can thus imagine that detecting changes in the neutronrefractive index is a rather ambitious goal. It has been estimated \cite{Havermeyer-apb99} that typical changes are in the order of $10^{-10}$. Therefore, a different approach was used to tackle the problem: The photo-neutronrefractive sample was illuminated with a sinusoidal light interference pattern thus achieving space-charge fields $|\vec{E}(\vec{x})|$ of up to 100 kV/cm, a magnitude which exceeds by far the values  usually achieved. As a consequence the neutronrefractive-index is modulated via the neutron electrooptic effect according to  (\ref{eq:neo}). Then neutrons are diffracted from these holographically recorded grating. So we are able to detect the neutronrefractive-index changes under the influence of (spatially modulated) electric fields with the extraordinary sensitivity required. 
\subsection{Methodology to determine the ENOC}
The experiment for the determination of the ENOC in $\rm LiNbO_3$ was performed in three steps: the recording of the grating by a two-wave mixing technique, measuring the rocking-curve for light, and subsequently for neutrons. In the first step the space-charge field is created, then its magnitude $E_{sc}$ is calculated according to  (\ref{eq:eo}), as we know the refractive index and the electrooptic coefficient for light. Then the sample was transferred to the D22 at Grenoble at cryogenic temperatures to prevent thermal decay of the space-charge field. From the measured angular dependence of the diffraction efficiency for neutrons $\eta_N(\theta)$ the neutronrefractive-index change was estimated and in turn the ENOC $r_N^T$ by  using  (\ref{eq:neo}). 

\begin{figure}
  \centering
  \apbgraph{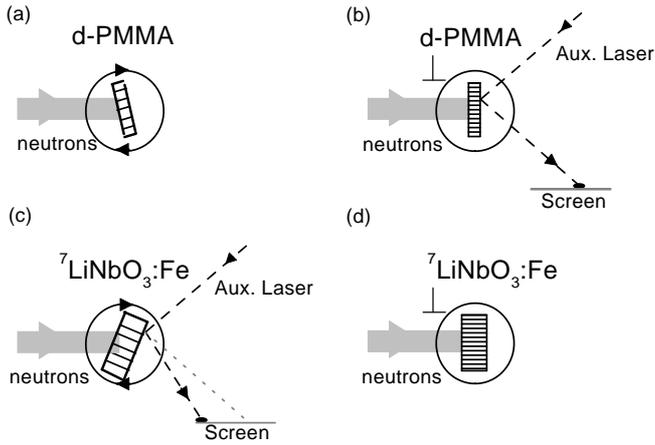}
  \caption{Adjustment procedure to find the Bragg reflection for determining the ENOC in \linbo. 
(a) Aligning a d-PMMA sample-surface perpendicular to the neutron beam. (b) Reflecting an auxiliary laser from the d-PMMA sample-surface to a screen. (c) Exchanging d-PMMA with the \linbo. (d) Re-adjusting the \linbo-surface to hit the initial reflection spot.
\label{fig:adjust}}
\end{figure}

This conceptually simple investigation goes hand in hand with quite many difficulties: 
      \begin{itemize}
\item Sample preparation: Without doubt the most studied photorefractive electrooptic material is $\rm LiNbO_3$:Fe. For this reason and its favourable properties (low conductivity and very long decay times at room temperature, mechanical and thermal stability, a mature technique in crystal growth and doping) we chose it to explore the neutron electrooptic effect. However, the natural abundance of $\rm Li$ consists of 7.5\% $\rm ^6Li$ with a $b_c=(2.15+ 0.26i)$ fm and 92.5\% $\rm ^7Li$ with a $b_c=-2.22$ fm. To avoid absorption, a monoisotopic \lisev-crystal was grown, on which the experiments were conducted. Moreover, we aimed at a fairly big crystal (several cm$^3$) for the neutron experiments. On the other hand, an increase in thickness usually gives rise to problems in recording, in particular as holographic scattering is enhanced  \cite{Stevendaal-josab98}. The high costs incurred did not allow for any more efforts and attempts at crystal growth and therefore the optical quality was not as good as we had intended it to be.
\item Owing to that, the determination of the space-charge field was not possible. It turned out that the grating had an effective thickness of about 1 mm despite the crystal's dimension of 12 mm. This led to a substantial reduction of the diffraction efficiency ($\eta\propto d^2$).
\item Another technical problem was, that for neutrons with a wavelength of $\lambda=1.4$ nm and a typical grating spacing in the order of $\Lambda=400 $ nm, Bragg angles around $\Theta_B=0.1$ deg result. Using the estimation from Ref.  \cite{Havermeyer-apb99} for $r_N$ and a space-charge field of 100 kV/cm we expect maximum diffraction efficiencies of $\eta_N\approx 10^{-5}$. This means that we have to search for a diffraction signal of $1/10^5$ in an angular range of 0.2 deg! Moreover, the flux for cold neutron wavelengths and good collimation ($x_n$=16 mm, $x_s$=9 mm, $L_c$=19.1 m) reduces the counting rate considerably. Therefore, a sophisticated adjustment procedure was applied: By measuring neutron rocking-curves of a d-PMMA sample, we determined the relative position of the neutron beam to the surface-normal. By means of an auxiliary laser the light reflection from the surface at a far distance was marked, and the d-PMMA exchanged for the \linbo-sample. Finally, the latter was adjusted to hit the same reflection spot. The orientation of the grating with respect to the sample surface had been determined earlier by measuring rocking-curves (for light). This procedure proved accurate enough to identify the very weak Bragg reflection (Sketch in Figure  \ref{fig:adjust}). 
      \end{itemize}
      \subsection{Experiment}
\begin{figure}
  \centering
  \apbgraph{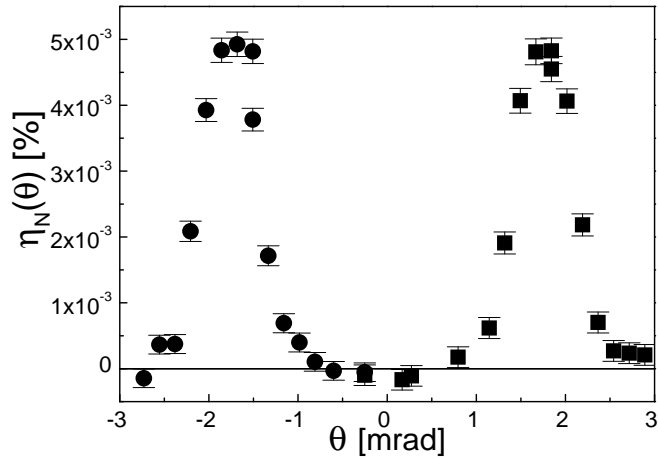}
  \caption{Angular dependence of the diffraction efficiency for cold neutrons. The latter are diffracted from a grating induced by the electro neutron-optic effect in \lisev. +1. and -1. diffraction orders are visible.\label{fig:enoc}}
\end{figure}

\begin{figure}
  \centering
  \apbgraph{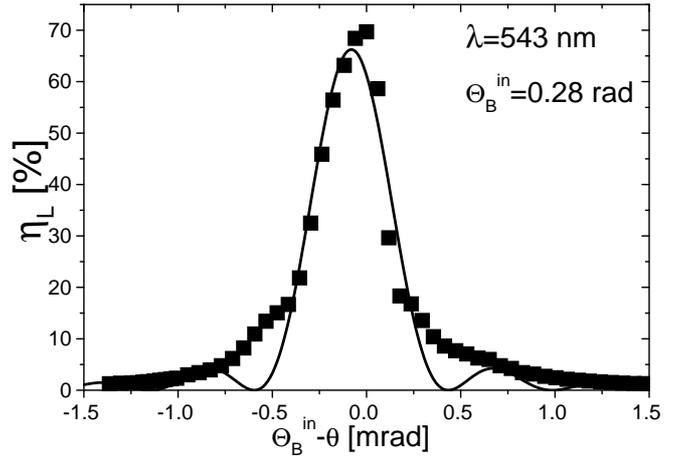}
  \caption{Light rocking-curve to determine the space-charge field ($\lambda=543$ nm). The solid line represents a fit to the data according to  (\ref{eq:diffractionefficiency}). \label{fig:enoclight}}
\end{figure}

\begin{figure}
  \centering
  \apbgraph{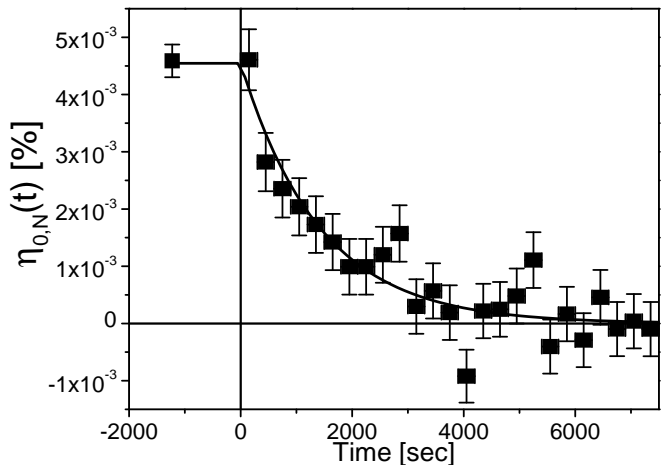}
  \caption{$\eta_0(t)$ during illumination with white light, $t=0$ indicates the start of the illumination. Counts were collected for 20 min, the data points are an average over this time span. \label{fig:enoctime}}
\end{figure}

The experiment to search for a neutron electrooptic effect was suggested by Rupp. In Ref. \cite{Rupp-om95} he gave an overview on tentative standard (light) electrooptic materials including estimations for the magnitudes of their ENOCs. A few years later the first experimental evidence of the effect was reported \cite{Havermeyer-apb99}. Gratings with a spacing of $\Lambda=389$ nm were recorded in a \lisev~ of dimensions $22\times 13 \times 12$mm$^3$. Then ten angular positions with cold (unpolarised) neutrons with a wavelength of $\lambda=1.2$ nm were probed to match the Bragg angle at the SANS facility D11 at the ILL. For each setting the counting time was 1 h. For one of the positions a Bragg peak was visible. However, because of lateral divergence ($x_n=15$ mm, $x_s=7$ mm, $L_c=40.5$ m) a complete rocking-curve has to be measured to extract the neutronrefractive-index change (\ref{eq:BWGint}) and finally the ENOC using  (\ref{eq:neo}). This task was performed in a subsequent experiment using the same sample but instrument D22 \cite{Rupp-osa99}. The grating spacing recorded and determined by means of light optics this time was $\Lambda=413\pm 2$ nm, the neutron wavelength $\lambda=1.39$ nm, $\Delta\lambda/\lambda\approx 10$\%, and the lateral divergence less than 1.3 mrad. For appropriate adjustment of the \lisev~ we employed the method described above and succeeded in measuring a neutron rocking-curve as can be seen in Figure  \ref{fig:enoc}. 
Again neutrons at each of the angular positions were accumulated for 1 h. Note that the maximum diffraction efficiency $\eta_0=5\times 10^{-5}$ is 4 to 5 orders of magnitude lower in contrast to typical ones for d-PMMA samples. Though in principle those measurements are sufficient to determine the ENOC, an unexpected problem occurred. Due to the bad optical quality of the sample and its large thickness, we were not able to extract the grating thickness and hence the magnitude of the space-charge field from the light optic rocking-curve shown in Figure  \ref{fig:enoclight}. We suppose that the recorded refractive-index pattern had a strong profile along the thickness direction. Fitting Kogelnik's equation to the light optic rocking-curve, we arrived at $d\approx 1.2$ mm and $\nu\approx 0.8$. Those values should be considered crude estimations only as obviously the experimentally measured rocking-curve could obviously not simply be described by the Kogelnik formula \cite{Kogelnik-Bell69}(see solid line in Figure  \ref{fig:enoclight}). 
The difference between the sample thickness and the grating thickness can be explained by extremely efficient holographic scattering so that the recording beams are already depleted within a few microns below the surface of the crystal. A new experiment with a thin crystal to evade these problems is under preparation. For the aforementioned reasons it is still not possible to give the exact value for the ENOC of \lisev. However, based on the data presented above, we estimate the ENOC to be $r_N^T\approx 3\pm 2$ fm/V. To prove that the gratings originate from a light-induced space-charge field and the electro neutron-optic effect,  the diffraction efficiency $\eta_0(t)$ at the Bragg position was measured during illumination with white light. The resulting time dependence is depicted in Figure  \ref{fig:enoctime}.
The decay of the diffraction efficiency gives evidence that the charge carriers composing the space-charge field are electrons and that the observed effect is indeed due to the coupling of the neutronrefractive-index change to that field via electro neutron-optics (cf.  (\ref{eq:neo})). 
\subsection{Application of the photo-neutronrefractive effect}
Up to this point one might assume that the electro neutron-optic effect is of pure fundamental interest. However, it will be shown that the method proposed can be used - quite on the contrary - to solve applied physical questions as well.
\subsubsection{Thermal fixing in hydrated and dehydrated \linbo:{\rm Fe}}
Apart from the advantages of holographic data storage using photorefractive crystals, several problems that are still an obstacle for commercial use, still need to be solved. One such critical element is that holographic memory is basically volatile, i.e. after recording the data with light, they are erased upon homogeneous illumination. The decisive physical parameter for this phenomenon is photoconductivity. Even worse, the data decay until complete disappearance even without exposure to light as a consequence of dark conductivity of the photorefractive material. Therefore, it is evident that fixing mechanisms have to be developed for long-term data storage. Several attempts to solve the problem are reported in literature
 \cite{Muller-jap95,Buse-n98,Wesner-apb01,Shen-apl00}. Among them the most common technique for \linbo, the material in question, is called \textit{thermal fixing} \cite{Muller-jap95,Yariv-josab96,Buse-prb97,An-ao99,Nee-jap00,Yang-apl01}. The underlying idea is as follows: As discussed in section \ref{sec:linbo}, holographically recording a grating means, that a light-induced space-charge density of electrons modulates the refractive index via the electrooptic effect. By increasing the temperature to about 400-450 K, positively charged ions become mobile and neutralise the effect of the electronic space-charge pattern, i.e. the (light) refractive-index grating disappears ($E_{sc}=0$). However, a density grating of ions (and electrons) remains. After returning to ambient temperatures, homogeneous illumination redistributes the electrons. The ions that are insensitive to illumination because of their mass yield an ionic space-charge density and a refractive-index grating via the electrooptic effect, which exactly mimics the primary grating: it is fixed. This technologically important process depends significantly on the species of ions, the temperature dependence of their mobility, on their concentration and various other parameters. Spectroscopic investigations have revealed that the ionic density grating can usually be traced back to hydrogen ions \cite{Vormann-ssc81,Buse-prb97}. Surprisingly, thermal fixing also works if crystals are dehydrated. In the latter case the question arises which ion will be responsible for the compensation process. Employing the knowledge gained from measuring the ENOC, neutron diffraction from two different \linbo-samples (hydrated, dehydrated) in two states ($E_{sc}=0, E_{sc}\not=0$) was performed to determine the species of ions \cite{Nee-prb99}.
\subsubsection{Experiment}
The experiment was performed in a similar way in case of the previously described measurement of the ENOC in \lisev: The instrument D22 at the ILL was used with the corresponding values for wavelength, collimation and counting rates. The two iron-doped \linbo~ samples had a thickness of 2.9 mm with natural isotopic composition of Li, i.e. absorption was enhanced considerably. Holographic gratings were prepared at 450 K, with a spacing of $\Lambda=374$ nm. The concentration of OH$^-$-ions, obtained via absorption measurements, was $5.64\times 10^{24}$ m$^{-3}$ and $0.47\times 10^{24}$ m$^{-3}$ in sample $\rm S_{hyd}$ and $\rm S_{dehyd}$ respectively. Rocking curves for both samples were conducted in each of the states. Due to lack of measurement time not all of them could be completed.

\begin{figure}
  \centering
  \apbgraph{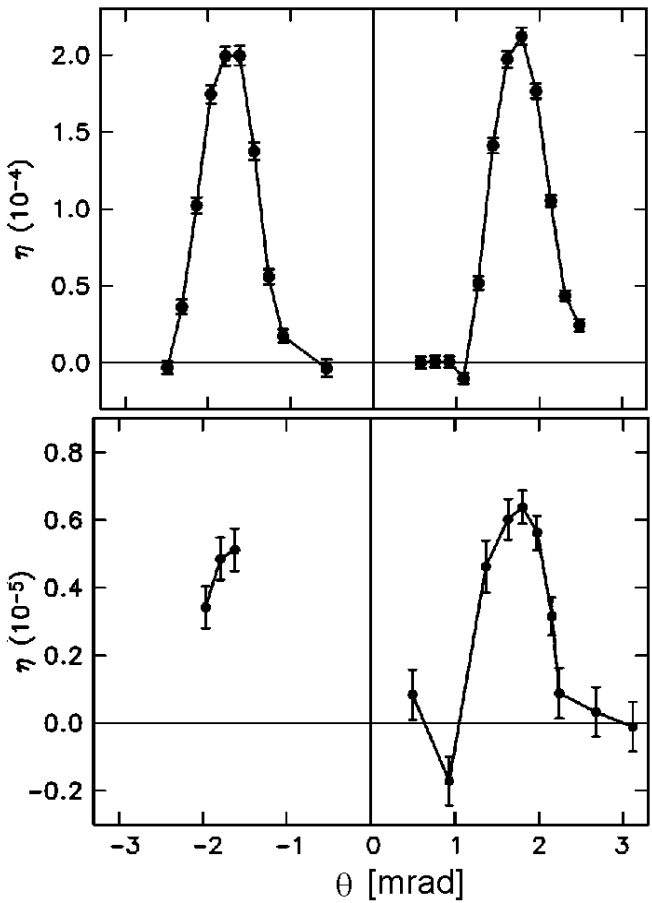}
  \caption{Rocking curve performed on the hydrated sample. Diffraction from the pure ionic density grating (bottom graph) and from the combined grating (top) \cite{Nee-prb99}. Lines are guides to the eye. \label{fig:thermfixh}}
\end{figure}

The conclusions which can be drawn from the four measurements are the following:
If $E_{sc}=0$, then the photo-neutronrefractive effect is only due to ionic density gratings, i.e. $\Delta n_{ion}=-\lambda^2/(2\pi))b_c\Delta c_{ion}$. For the hydrated sample we supposed that the whole contribution originates from hydrogen ions, for the dehydrated sample from the unknown ion species. If $E_{sc}\not=0$, the photo-neutronrefractive effect consists of two terms, the ionic density grating as before and the electro neutron-optic (ENO) part according to  (\ref{eq:neo}), $\Delta n=\Delta n_{ion}+\Delta n_{ENO}$.
Evaluating the neutronrefractive-index changes on the basis of the rocking-curves measured and as we know the concentration of ions, the coherent scattering length of the ion species responsible for the diffraction from the density grating can be calculated. In Figure  \ref{fig:thermfixh} the angular dependence of the diffraction efficiency for the hydrated sample is shown.
The evaluation for sample $\rm S_{hyd}$ in case of $E_{sc}=0$ yields $\Delta n_{ion}=9.2\times 10^{-10}$, which equals a $|b_c|=3.6$ fm for the compensating ion. This ties in nicely with the assumption that hydrogen ($b_c=-3.74$ fm) is responsible for the compensation process. Moreover, by adding or subtracting $\Delta n_{ion}$ from the total neutronrefractive-index change as evaluated for $E_{sc}\not=0$, the ENOC for \linbo~ can be estimated as $r_N=(2.4-3.4)\pm0.5$ fm/V. Equivalent measurements and evaluations for the sample $\rm S_{dehyd}$ have shown that any other effect is more than an order of magnitude smaller. Therefore, only the upper limit for $\Delta n_{ion}<6.5\times10^{-10}$ can be estimated, which in turn yields $|b_c|\leq 1.92$ for the compensating ion. This indicates that thermal fixing in dehydrated \linbo~ may be attributed to the motion of $\rm Li$-ions ($b_c=-1.9$ fm).
\section{Outlook and Summary\label{sec:discussion}}
Some of the novel aspects opened up by the use of photo-neutronrefractive samples have already been discussed in the relevant previous sections (see e.g. section \ref{sec:nif}). Here, we will briefly suggest additional future experiments which promise great potential for cold neutron physics and materials science. Moreover, new tentative photo-neutronrefractive materials are briefly discussed.
\subsection{Future experiments}
A straightforward experiment to be conducted is an Aharonov-Casher (AC) type diffraction experiment \cite{Rupp-om95}. Using the same experimental configuration as in the determination of the ENOC but employing polarised neutrons, the value of $r^S$ according to  (\ref{eq:Schwinger}) and (\ref{eq:enocS}) will depend on the mutual orientation of the neutron spin, the space-charge field, and the propagation direction. One of the advantages in our setup is that $V_S$ is enhanced by the factor $\varepsilon=30$ in \linbo. Moreover, by choosing an appropriate geometry, i.e. $\vec{E}\times\vec{k}$ parallel or antiparallel $\mu$, we can make use of a heterodyne technique: The ENOC for unpolarised neutrons will be enhanced or diminished. As the effect is quite small, there is need for an improved accuracy. The latter can be achieved by periodically flipping the spin state (e.g. every 15 min) and finally summing up the corresponding counts. Thus any systematic drift or fluctuation will cancel out. Note that integration times of one hour at the D22 (ILL) were necessary to observe the electro neutron-optic effect with an acceptable accuracy using unpolarised neutrons. The proposed experiment has to be performed on a high flux source at a SANS facility like the D22. However, up to now polarised neutrons are not available there. Attempts at the PSI failed because of the low flux for cold neutron wavelengths employing good collimation. Using  (\ref{eq:enocS}) and (\ref{eq:neo}) the AC contribution of the ENOC to the diffraction efficiency is
    \begin{equation}\label{eq:ac}
  \eta_{(AC)}=\left(\frac{|\vec{\mu}|\pi\varepsilon E d}{hc}\right)^2,
    \end{equation}
and thus independent of the neutron wavelength $\lambda$ and the grating spacing $\Lambda$. Moreover, for further experiments it is noteworthy that the AC effect is suppressed by choosing $\vec{\mu}\parallel\vec{E}$.

An interesting aspect of the Foldy-contribution (second term in (\ref{eq:enocS})) is its appearance as a non-local effect. The corresponding neutronrefractive-index modulation is phase-shifted by $\pi/2$ with respect to that of the density modulation (\ref{eq:enocT}). Thus, measuring the phase shift, e.g. by means of an interferometric technique, it might be possible to determine its contribution to the ENOC. Because of the small diffraction angles, the Foldy-contribution, however, is three to four orders of magnitude smaller than the AC-effect. 

The measurement of the ENOC also provides a possibility for detecting an EDM of the neutron. Novel techniques have been developed, suggested and employed for this purpose \cite{Alexandrov-92,Fedorov-phb97,Zeyen-nima00,Harris-nima00}.
Under illumination huge electric space-charge fields build up in the electro neutron-optic crystal which can not be reached in vacuum. A combination of the interferometric technique and the use of those electric fields is a tentative method to lower the limit for the existence of an EDM.
\subsection{Further promising photo-neutronrefractive materials}
At first we consider materials where the neutron-optical potential is governed by the nuclear contribution so that  (\ref{eq:nopt}) is valid. Then, in general, we can distinguish two possibilities to modulate the potential: either changing the density $\Delta\rho$ as already discussed or the mean bound scattering length $\overline{\Delta b}$, which would mean that light-induced isotope separation occurs. It is evident that it is much more realistic to change the density. We suppose that any of the photosensitive polymers, which nowadays are available in large numbers, with a considerable ratio $\rho_{polymer}/\rho_{monomer}$, is suitable as effective photo-neutronrefractive material (e.g.  \cite{Eickmans-jjapa99,Zilker-apb99,Steckman-TOPS00,Bieringer-TOPS00,Semenova-spie01}). An attempt towards using another photopolymer was also made, employing poly(methyl-2-cyanoacrylate), PMCA \cite{Rupp-p97}. In this material, a different polymerisation mechanism (anionic polymerisation) as compared to PMMA may be active. However, it has turned out that the characteristics are very similar. Performing neutron diffraction on a protonated sample, the maximum diffraction efficiency corrected for absorption amounted to about $\eta_0=10^{-3}$. Due to the fact that $\overline{b}$ for PMCA is higher than for PMMA, a deuterated sample is promising material. Preliminary research on another protonated system, PMMA/phenanthrenequinone, has already been per\-formed \cite{Pruner-98}. This material exhibits diffuse amplification and can be used as an extremely thick holographic recording medium \cite{Popov-joa00}. The neutron diffraction experiments showed a diffraction efficiency $\eta_0\approx3\times10^{-4}$ at the Bragg-angle.

Considering materials which are photo-neutronrefractive via the electro neutron-optic effect, one must admit that the effects will be small when compared to the photopolymer systems. However, the advantage of these materials is that they are well established in light optics and respond with linear neutronrefractive-index changes to illumination. Moreover, the magnitude of the refractive-index changes depends on the ENOC and the space-charge field. Estimations for various common light electrooptic materials are included in Ref.  \cite{Rupp-om95}.

Similar to thermal fixing in \linbo, ionic (=density) gratings may be created in any photorefractive sample with high ionic conductivity. Promising material for this type of effect is $\rm LiIO_3$, which is known as a quasi onedimensional ionic conductor at ambient temperatures. Moreover, photorefractive properties have been proven to exist at temperatures below 180 K \cite{Xu-prb96,Xu-prb98,Sun-oc01}. As the neutronrefractive-index changes are proportional to the concentration modulation $\Delta c_{ion}$ of the specific ion, we expect that at lower temperatures the ionic grating must be very efficient in diffracting neutrons. A first attempt with iron-doped $\alpha$-$\rm LiIO_3$ using  the HOLONS facility and a cryostat at low temperatures failed, however. Again the huge absorption cross section of ~$\rm ^6$Li seemed to be the major obstacle.

Another very interesting possibility to observe light-induced neutronrefractive-index changes is the use of photomagnetic samples and polarised neutrons, which was suggested by Rupp  \cite{Rupp-WS01}. Only recently have a photorefractive effect and light-induced birefringence been observed in garnets \cite{Sugg-apl95,Sugg-om95,Sugg-josab96,Chen-apb98}, in $\rm Tb_3Ga_5O_{12}$ even at room-temperature. 
Moreover, the fact that the bound scattering length in general is spin dependent will be of utmost importance. Provided that the nuclear spins are or can be (re)oriented, extremely efficient diffraction of polarised neutrons from light-induced gratings can be expected. Transparent magnetic borates, e.g. $\rm FeBO_3$, are also highly promising materials.
\subsection{Future Devices}
The development of useful devices for neutron optics is based on the simplest optical element: the grating. The combination of gratings and/or the modification of the light-optical setup allows us to prepare new devices. A single grating can be used already as a monochromator.
By creating more complex neutronrefractive-index profiles, e.g. lenses for cold and ultracold neutrons can be designed. Only recently did Oku perform this task for cold neutrons based on compound refractive Fresnel lenses, which consisted of about 50 elements \cite{Oku-phb00,Oku-nima01}. We are convinced that the fabrication of such a lens by holographic means is much simpler and cheaper. Because of the lateral divergence of the neutrons, part of the beam cannot be in exact Bragg-position. To efficiently reflect all neutrons of the beam by Bragg-diffraction from gratings, we are planning to record multiple gratings with slight rotations between the corresponding grating vectors. This will lend great assistance in fabricating mirrors. However, work towards these directions is still in its infancy. 

To summarise, a range of phenomena related to light-induced changes of the refractive index for neutrons was presented. This photo-neutronrefractive effect was defined in terms analogous to light optics. The preparation of gratings based on this effect by light-optical means and the diffraction of cold neutrons from that gratings were discussed. We showed that combined neutron and light-optical experiments yield important information on material properties. This knowledge in turn is useful for producing neutron-optical elements. The design, setup and the successful operation of a neutron interferometer based on such holographically recorded gratings was presented together with first results. 

By diffracting cold neutrons from gratings in substances exhibiting the electro neutron-optic effect, i.e. by changing the neutronrefractive index under the application of electric fields, fundamental properties of the neutron are probed. Preliminary experiments and related results were presented, the article finishing with planned experiments and a discussion of their impact on the foundations of physics.
\begin{acknowledgement}
I would like to express my gratitude to Romano A. Rupp for drawing my attention to this exciting topic and appreciate very much his enthusiastic and stimulating way of discussing open problems. I am indebted to Frank Havermeyer and Christian Pruner for providing me with unpublished data. Gerhard Krexner, Romano A. Rupp, Christian Pruner and Romana Zeilinger critically read the manuscript. To them I offer my sincere thanks.

Financial  support by the grant FWF P-14614-PHY and the Austrian ministry bm:bwk (infrastructure for HOLONS at the GeNF) is acknowledged. Particular thanks go to the crew at the GeNF for their manifold activities to set HOLONS into operation.
\end{acknowledgement} 
\bibliographystyle{apbsty} 

\end{document}